\documentclass[english,12pt,aps,prd,a4paper,preprintnumbers,floatfix,nofootinbib,showpacs,superscriptaddress, notitlepage]{revtex4-1}
\pdfoutput=1

\usepackage[english]{babel}
\usepackage{amsmath}
\usepackage{graphicx}
\usepackage{color}
\usepackage{mathrsfs}   
\usepackage{amssymb}
\usepackage{enumerate}
\usepackage[normalem]{ulem} 
\usepackage{dcolumn}
\usepackage{bm}
\usepackage{graphicx}
\usepackage[sort&compress]{natbib}
\usepackage{epstopdf}
\usepackage{caption}
\usepackage{subcaption}
\usepackage{multirow}
\usepackage{rotating}
\usepackage[utf8]{inputenc} 
\usepackage{enumitem,xcolor}

\newcommand {\be} {\begin{equation}}
\newcommand {\ee} {\end{equation}}

\usepackage[colorlinks=true,citecolor=darkred,urlcolor=darkred, pdfborder={0 0 0}]{hyperref}
\usepackage[normalem]{ulem}

\definecolor{darkred}{rgb}{0.6,0,0}

\definecolor{linkcolor}{rgb}{0,0,0.5}

\definecolor{greenLinks}{rgb}{0, 0.6, 0} 
\definecolor{blueLinks}{rgb}{0, 0, 0.6}
\definecolor{redLinks}{rgb}{0.6, 0, 0}
\definecolor{tempText}{rgb}{0.55, 0.10,0.67}
\definecolor{eprintLinks}{rgb}{0.4, 0.4, 0.4}
\definecolor{journalLinks}{rgb}{0.6, 0, 0}

\usepackage{float}

\def\21{$\mathrm{SU(2)_L \otimes U(1)_Y}$ }
\def\31{$\mathrm{SU(3)_c \otimes U(1)_Q}$ }
\def\SM{$\mathrm{SU(3)_c \otimes SU(2)_L \otimes U(1)_Y}$ }

\def\3211{$\mathrm{SU(3) \otimes SU(2)_L \otimes U(1)_R \otimes U(1)_{B-L}}$ }
\def\321{$\mathrm{SU(3) \otimes SU(2) \otimes U(1)}$ }
\def\422{$\mathrm{SU(4) \otimes SU(2) \otimes SU(2)_R}$ }
\newcommand {\ignore}[1]{}

\newcommand{\sm}{{Standard Model }}

\def\SM{$\mathrm{ SU(3)_C \otimes SU(2)_L \otimes U(1)_Y }$ }

\newcommand{\AddrAHEP}{%
  AHEP Group, Institut de F\'{i}sica Corpuscular --
  CSIC-Universitat de Val\`{e}ncia, Parc Cient\'ific de Paterna.\\
 C/ Catedr\'atico Jos\'e Beltr\'an, 2 E-46980 Paterna (Valencia) - SPAIN}
\newcommand{\AddrUNAM}{ {\it Instituto de F\'{\i}sica, Universidad Nacional Aut\'onoma de M\'exico, A.P. 20-364, Ciudad de M\'exico 01000, M\'exico.}}
\newcommand{\AddrTUM}{ Physik-Department T30d, Technische Universit\"{a}t M\"{u}nchen.\\
James-Franck-Strasse, 85748 Garching, Germany}

\begin{document}

\bibliographystyle{unsrt}   

\title{The role of residual symmetries in dark matter stability and the neutrino nature}
\author{Cesar Bonilla}\email{cesar.bonilla@tum.de}
\affiliation{\AddrTUM}
\author{Eduardo Peinado}\email{epeinado@fisica.unam.mx}
\affiliation{\AddrUNAM}
\author{Rahul Srivastava}\email{rahulsri@ific.uv.es}
\affiliation{\AddrAHEP}

\begin{abstract}
  \vspace{1cm}
  
We consider the class of models where  Dirac neutrino masses at one loop and the dark matter stability
can be obtained using only the global $U(1)_{B-L}$ symmetry already present in Standard Model.
We discuss how the residual $\mathcal{Z}_n$ subgroup,  left unbroken after the breaking of $U(1)_{B-L}$, dictates the neutrino nature, namely if they are Dirac or Majorana particles, as well as determines the stability of the dark matter candidate in such models. In particular,  we show that without the correct breaking of $U(1)_{B-L}$ to an appropriate residual $\mathcal{Z}_n$ symmetry, the Dirac nature of neutrinos and/or dark matter stability might be lost. For completeness we also provide some examples where the dark matter stability is accidental or lost completely. Finally, we discuss one example model where the Dirac neutrinos with naturally small one loop masses as well as dark matter stability, are both protected by the same residual $\mathcal{Z}_6$ subgroup, without need for adding any new explicit or accidental symmetries beyond the  Standard Model symmetries.
\end{abstract}

\maketitle

\section{Introduction}
\label{sec:intro}

The Standard Model is the best description we have so far to explain 
all the observed fundamental particles and their interactions, namely the strong and 
electroweak phenomena. However, the \sm predicts that neutrinos are massless
particles and lacks a candidate to account for the dark matter relic abundance 
in the Universe~\cite{Aghanim:2018eyx}.

Neutrino oscillation experiments indicate that at most one active neutrino can be 
massless~\cite{Whitehead:2016xud, Decowski:2016axc, Abe:2017uxa, deSalas:2017kay}
but there is no experimental hint pointing towards the exact mechanism to generate
mass for neutrinos. In this regard, the most popular approach to alleviate this
\sm shortcoming is to assume that neutrinos are Majorana in nature and invoke the so-called Majorana $seesaw$ mechanisms~\cite{Minkowski:1977sc,Yanagida:1979as,Mohapatra:1979ia,Schechter:1980gr,Schechter:1981cv,Foot:1988aq}.
In fact, independent of the Dirac/Majorana nature of neutrinos and the details of mass generation mechanism, neutrino mass generation  always requires the existence of new particles and/or symmetries~\cite{Ma:1998dn, CentellesChulia:2018gwr}. 

From the theoretical perspective it is very tempting to think that one 
of the new fundamental fields added to the \sm to generate the neutrino mass might be 
a potential dark matter candidate. The connection between dark matter and the neutrino mass naturally arises when the mass generation mechanism is at the loop level. The simplest realization
of this idea is achieved within the $scotogenic$ model~\cite{Ma:2006km}. 
Besides the \sm particle content, in this scenario the neutrino mass is generated at the one-loop level by assuming the existence of \SM gauge singlet fermions  and an extra $SU(2)_L$ doublet scalar with vanishing vacuum expectation value (vev). All these new particles carry an odd charge under a global $\mathcal{Z}_2$-symmetry.
This additional $\mathcal{Z}_2$ symmetry  remains unbroken after the electroweak symmetry breaking (EWSB). It  stabilizes the dark matter candidate and forbids the appearance of Majorana mass terms for the neutrinos at tree-level.

Following the scotogenic idea, here we will consider a class of models~\cite{Bonilla:2018ynb} where\footnote{For analogous framework for Majorana neutrinos see~\cite{CentellesChulia:2019gic}. }
\begin{enumerate}
\item[ I.] Neutrinos are Dirac in nature.
\item[ II.]  Neutrino mass is generated at one loop level. 
\item[ III.] In scotogenic spirit, the intermediate particles in the loop belong to a ``dark sector'' with the lightest particle among them being a good candidate for stable dark matter.
\end{enumerate}
Typically, one needs to invoke several additional symmetries to achieve this. However, as has been shown in~\cite{Bonilla:2018ynb}, for certain models this can be achieve only through the global $U(1)_{B-L}$ of the \sm\footnote{The idea of obtaining dark matter stability associated to the breaking of $U(1)_{B-L}$ has  been explored in~\cite{Kadastik:2009dj,Kadastik:2009cu,Ma:2015xla, Chulia:2016ngi, Chulia:2016giq, Wang:2016lve,CentellesChulia:2017koy,Wang:2017mcy,CentellesChulia:2018gwr, Han:2018zcn,Lazarides:2018aev, Bonilla:2018ynb,Calle:2018ovc,CentellesChulia:2019gic}. For recent works aimed at obtaining scotogenic stability from extended gauge symmetries see~\cite{Ma:2019yfo,Kang:2019sab}. }.
In this setup the global $U(1)_{B-L}$ symmetry of \sm is broken down to its residual $\mathcal{Z}_n$; $n \in \mathbb{Z}^+$ subgroup. Since we require neutrinos to be Dirac particles, this residual symmetry should be such that it can protect the Dirac nature of neutrinos as well as stabilize the dark matter candidate~\cite{Bonilla:2018ynb}. This can be successfully done only if the breaking $U(1)_{B-L} \to \mathcal{Z}_n$ is achieved in a correct manner such that an appropriate $\mathcal{Z}_n$ subgroup is left unbroken. Not all $\mathcal{Z}_n$ subgroups of $U(1)_{B-L}$ can do this job. The criterion for such an appropriate residual $\mathcal{Z}_n$ are also listed down in~\cite{Bonilla:2018ynb, CentellesChulia:2019gic}.  In this paper we aim to construct a class of models and analyze them in details, in order to highlight the crucial role that the residual $\mathcal{Z}_n$ plays in protecting the Dirac nature of neutrinos as well as in dark matter stability.

This work is organized as follows: in the next section we describe how the residual symmetry is connected to the neutrino nature. In section~\ref{sec:one-loop} the list of possible realizations to generate the neutrino mass dynamically is given. In order to highlight the important role of the residual symmetry in ensuring the dark matter stability, we provide  several explicit examples in Section~\ref{sec:symmetryrole}, where, due to lack of an appropriate residual symmetry, the dark matter stability is lost. For completeness we also give examples of the cases where an additional accidental symmetry might be present in the model leading to dark matter stability. In Section~\ref{sec:model} we present a model where the residual $\mathcal{Z}_6$ symmetry  protects both the stability of the dark matter candidate and Dirac nature of neutrinos. We conclude after that.


\section{Residual symmetries and the neutrino nature}
\label{sec:dir-maj}


Typically in high energy physics models, when a symmetry $G$ is broken down either spontaneously or explicitly, a subgroup of it remains unbroken. This unbroken 
subgroup $G_{\text{res}}$ we refer to as the residual subgroup. It is this residual subgroup $G_{\text{res}}$ and not the full group $G$, which dictates the dynamics of a given theory at energy and temperatures below the scale of $G$ breaking. The most well known example of this is provided by \sm itself.
In \sm the gauge group \SM is spontaneously broken down to a residual subgroup $G_{\text{res}} \equiv \mathrm{SU(3)_c \otimes U(1)_{em}}$ by the vev of the Higgs.
Below the electroweak breaking scale, the dynamics of the \sm is dictated by the unbroken residual subgroup $\mathrm{SU(3)_c \otimes U(1)_{em}}$. 
Apart from this well known example, residual symmetries occur in almost all beyond \sm (BSM) extensions, with \SM itself being a residual subgroup in most of the BSM gauge extensions. Coming back to neutrinos, the residual symmetries play an important role in determining the Dirac/Majorana nature of neutrinos and might also be responsible for dark matter stability. We first highlight their role in determining the nature of neutrinos before discussing their role in dark matter stability.  

 To see the important role played by residual symmetries in determining the nature of neutrinos, recall that apart from \SM gauge symmetries, the \sm also has two other symmetries namely Lepton number $U(1)_L$ and Baryon number $U(1)_B$ symmetries.  The combination $U(1)_{B-L}$ can be rendered anomaly free if we extend the \sm particle content by adding three right handed neutrinos $\nu_{R_i}$; $i=1,2,3$ either carrying vector charges $(-1,-1,-1)$ or chiral charges $(-4,-4,5)$ 
under $U(1)_{B-L}$ symmetry~\cite{Montero:2007cd,Ma:2014qra,Ma:2015mjd,Ma:2015raa}. Given that the \sm lepton doublets $L_i$; $i =1,2,3$ carry $U(1)_{B-L}$ charge,
the  Dirac or Majorana nature of neutrinos depends crucially on the residual symmetry $G_{\text{res}}$ appearing after the breaking of the $U(1)_{B-L}$ symmetry. Since $U(1)_{B-L}$ is an Abelian continuous group, it only has discrete Abelian subgroups, namely $G_{\text{res}}\equiv\mathcal{Z}_m$; $m \in \mathbb{Z}^+$. Depending on the nature of the residual $\mathcal{Z}_m$ subgroup left unbroken,  one can classify the Dirac or Majorana nature of neutrinos, depending on how \sm lepton doublets $L_i$; $i = 1,2,3$ transform under the residual subgroup\footnote{If there are new, as of yet unknown, conserved symmetries present in nature, the classification presented here should be generalized to include them. Such a generalization is straightforward.}. If $U(1)_{B-L}$ remains unbroken then neutrinos will be Dirac particles as the Majorana mass term for neutrinos is forbidden by $U(1)_{B-L}$ symmetry. In case $U(1)_{B-L} \to \mathcal{Z}_m$ we have~\cite{Hirsch:2017col}.
\begin{eqnarray}
U(1)_{B-L}   & \, \to  \, &   \mathcal{Z}_m \equiv \mathcal{Z}_{2n+1} \, \text{with} \,  n \in \mathbb{Z}^+    \nonumber \\
& \, \Rightarrow \, & \text{Neutrinos are Dirac particles}      \nonumber \\
\label{eq:oddzn}
U(1)_{B-L}  & \, \to  \,  & \mathcal{Z}_m \equiv \mathcal{Z}_{2n} \, \text{with} \,  n \in \mathbb{Z}^+  \\
& \, \Rightarrow \, & \text{Neutrinos can be Dirac or Majorana} \nonumber
\end{eqnarray}
Now, if the $U(1)_{B-L}$ is broken to a $\mathcal{Z}_{2n}$ subgroup, then one can make 
a further classification depending on how the $L_i$ transform,
\begin{eqnarray}
 L_i \left\{ \begin{array}{ll}
          \nsim  \omega^{n} \ \ \text{under $\mathcal{Z}_{2n}$} &  \Rightarrow\text{Dirac Neutrinos}\\
          \sim  \omega^{n}\ \ \text{under $\mathcal{Z}_{2n}$} &  \Rightarrow\text{Majorana Neutrinos}\end{array} \right. 
\label{evenzndir}
\end{eqnarray}
where $\omega = e^{2 \pi I/2n} ;\omega^{2n}=1$ is the $2n$-th root of unity.

In order to understand better our last statement, one can take, for instance,
the dimension-5 Weinberg operator~\cite{Weinberg:1980bf} for Majorana neutrino mass $\bar{L}^c L H H$ 
(where $L$ and $H$ are  $SU(2)_L$ lepton and scalar Higgs doublets, respectively) which 
basically breaks lepton number leaving a  residual $\mathcal{Z}_2$ symmetry unbroken.
Neutrinos will be Majorana in this case as it satisfies the criterion for Majorana neutrinos, listed in \eqref{evenzndir}. 
As an example of a ultra violet (UV) completion, one can take the type-I seesaw mechanism, where three RH-neutrinos are added to the \sm, transforming as $(\nu_{R_i})\sim(-1)$ (with $i=1,2,3$) under the
global $U(1)_{B-L}$ symmetry. In order to generate the Majorana mass term for neutrinos, a lepton number breaking term  leading to $U(1)_{B-L} \to \mathcal{Z}_2$ has to be added to the Lagrangian. Hence,
\begin{equation}
\mathcal{L_\nu}=y^{\nu} \bar{L} \tilde{H} \nu_R + M_R \bar{\nu}^c_R \nu_R +h.c.
\end{equation}
where we have omitted the flavor indices for convenience. In this case the mass term $ M_R \bar{\nu}^c_R \nu_R $ softly breaks the $U(1)_{B-L} \to \mathcal{Z}_2$ and hence neutrinos are Majorana in nature.
On the other hand, if a dynamical origin of the neutrino mass is demanded, one has to further assume 
the existence of a scalar singlet $\sigma$ with non-trivial charge under  $U(1)_{B-L}$, i.e. $\sigma \sim \pm 2$ should be chosen.
As consequence, the vev of this scalar $\langle\sigma \rangle$ will break $U(1)_{B-L} \to \mathcal{Z}_2$   ultimately generating the Majorana mass for neutrinos~\cite{Mohapatra:1979ia,Schechter:1981cv}. 

To obtain Dirac neutrinos one must break  $U(1)_{B-L} \to \mathcal{Z}_n$ with $n \geq 3$ as the residual subgroup  $\mathcal{Z}_2$ always leads to Majorana neutrinos in accordance with \eqref{evenzndir}. While traditionally mass models for Majorana neutrinos have garnered lot of attention, in recent years Dirac neutrino mass models are enjoying a resurgence of sorts with various seesaw~\cite{Ma:2014qra,Ma:2015mjd,Ma:2015raa,Valle:2016kyz,Chulia:2016ngi, Chulia:2016giq, Reig:2016ewy, CentellesChulia:2017koy, CentellesChulia:2017sgj, Borah:2017leo, Bonilla:2017ekt, Srivastava:2017sno, Borah:2017dmk, Borah:2018nvu}
and loops models~\cite{Farzan:2012sa,Okada:2014vla,Bonilla:2016diq, Wang:2016lve,Ma:2017kgb,Wang:2017mcy,Helo:2018bgb,Reig:2018mdk,Han:2018zcn,Kang:2018lyy,Bonilla:2018ynb,Calle:2018ovc,Carvajal:2018ohk,Ma:2019yfo,Bolton:2019bou,Saad:2019bqf} 
considered in literature\footnote{
For Dirac neutrino models without mass mechanism see~\cite{Aranda:2013gga,Heeck:2013rpa,Abbas:2013uqh,Abbas:2016qbl,Hirsch:2017col}.}. 
The operator based classification of such models at dimension-4~\cite{Ma:2016mwh}, dimension-5~\cite{Yao:2018ekp,CentellesChulia:2018gwr} and dimension-6~\cite{Yao:2017vtm,CentellesChulia:2018bkz} have also been considered.

From now on, our focus is to explain the smallness of the neutrino mass when neutrinos are Dirac particles in nature. That is, there should be a dynamical mass mechanism behind the Dirac neutrino mass generation.  In order to have a natural explanation for smallness of Dirac neutrino masses, it is desirable to 
forbid the tree-level neutrino Yukawa coupling $\mathcal{L}_{\nu}\supset y^{\nu} \bar{L} \tilde{H} \nu_R$. This coupling can be forbidden in many ways, for example by imposing flavor symmetries~\cite{Chulia:2016giq,CentellesChulia:2017koy} or even by a simple $\mathcal{Z}_2$ symmetry~\cite{Chulia:2016ngi}. 
It can be automatically forbidden by the $U(1)_{B-L}$ symmetry itself, if we use the chiral solution of $(\nu_{R_a},\nu_{R_3})\sim(-4,5)$ (with $a=1,2$), as was shown for the first time in~\cite{Ma:2014qra,Ma:2015mjd,Ma:2015raa}.
The operator that generates the neutrino mass will then appear at higher dimensional level. For this to happen, a singlet scalar field $\chi \sim 3$ under the $U(1)_{B-L}$ has to be added in order to generate the mass of at least two neutrinos through a dimension 5 operator. The dimension 5 operator then is given by
\begin{eqnarray}
 \mathcal{L}_\nu \supset \frac{y^{\nu}}{\Lambda} \bar{L} \, \tilde{H} \, \chi \, \nu_{R_a}
 \label{op}
\end{eqnarray}
where $\Lambda$ is the scale of UV completion. The mass of the third neutrino can also be generated by either the dimension-6 term
\begin{eqnarray}
 \mathcal{L}_\nu \supset \frac{y^{\nu}}{\Lambda^2} \bar{L} \, \tilde{H} \, \chi^* \, \chi^* \, \nu_{R_3}
\end{eqnarray}
or by adding a new singlet scalar $\chi_6 \sim 6$ under $U(1)_{B-L}$ such that the dimension-5 term
\begin{eqnarray}
 \mathcal{L}_\nu \supset \frac{y^{\nu}}{\Lambda} \bar{L} \, \tilde{H} \, \chi_6^* \, \nu_{R_3}
\end{eqnarray}
is also allowed by the $U(1)_{B-L}$ symmetry. 

The UV completion of such operators is model dependent, as we will discuss in Section~\ref{sec:one-loop}. 
Since the field $\chi \sim 3$ under $U(1)_{B-L}$, its vev will break $U(1)_{B-L} \to \mathcal{Z}_{3m}$; $m \in \mathbb{Z}^+$. The exact residual $\mathcal{Z}_{3m}$ subgroup will depend on the details on the UV completion, a fact we discuss in further details in Section~\ref{sec:symmetryrole}. Also, note that, since the $U(1)_{B-L}$ charge of $\chi_6 \sim 6$ is a multiple of the charge of $\chi \sim 3$, the addition of $\chi_6$ in a given UV complete model will not change the nature of the residual  
$\mathcal{Z}_{3m}$ symmetry. Before moving on let us briefly see the various possible ways the operator in \eqref{op} can be UV completed at one-loop.

\section{One-loop topologies of the operator $\bar{L} \, \tilde{H} \, \chi \, \nu_{R_a}$}
\label{sec:one-loop}%
\begin{figure}[H]
    \centering
    \begin{subfigure}[t]{\linewidth}
        \includegraphics[width=1.03\linewidth]{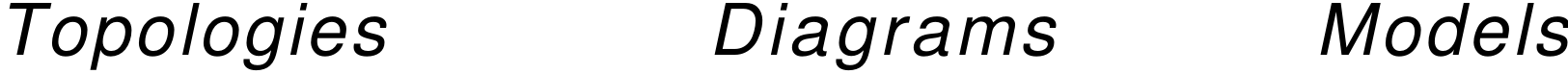}
    \end{subfigure}
       ~ ~ 
      \par\vspace{-.35cm}  
    \begin{subfigure}[b]{\linewidth}
   \includegraphics[width=\linewidth]{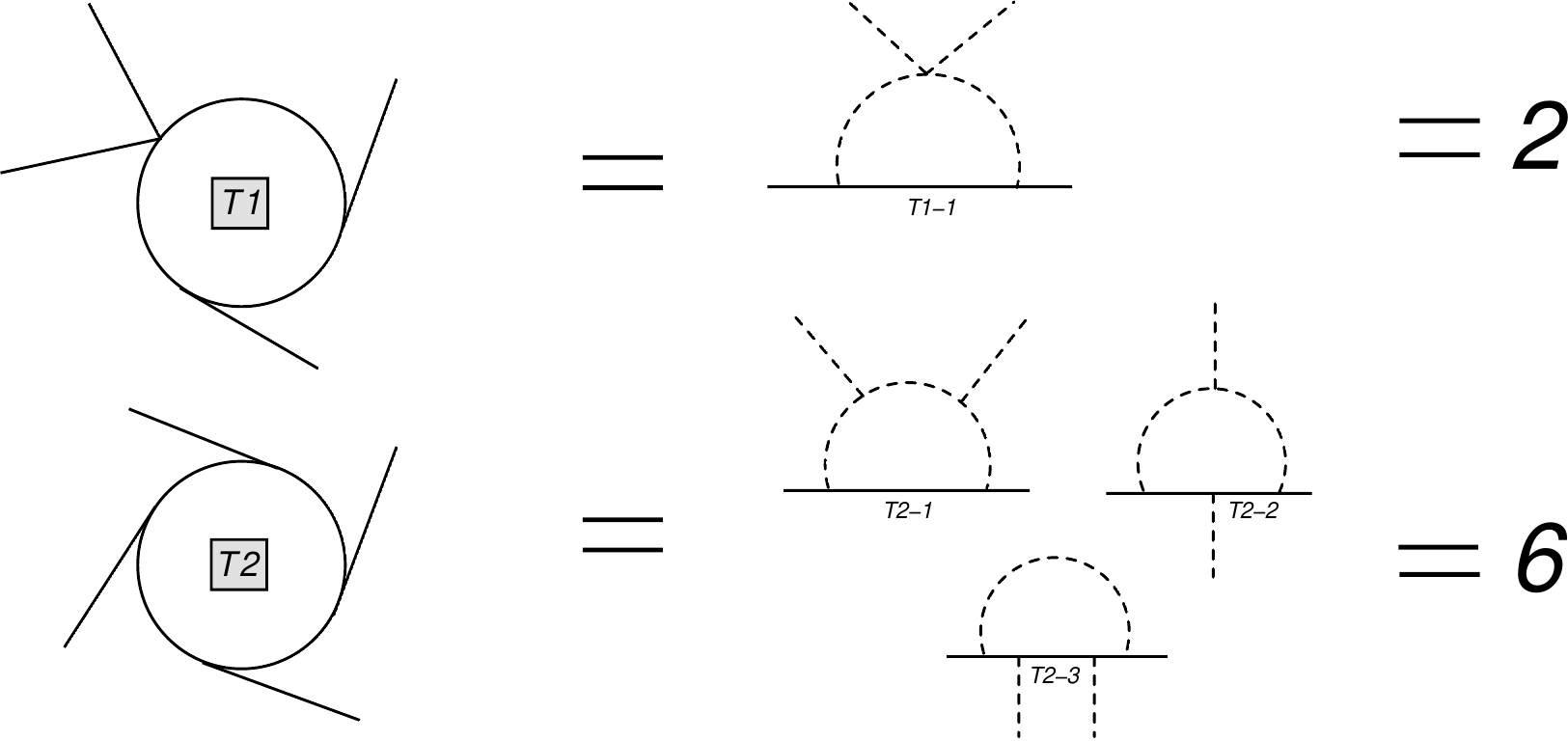}      \end{subfigure}
    ~ 
    \caption{One loop topologies with four external lines.}\label{fig:gencase}
\end{figure}

In this section we provide the topologies and all possible one-loop diagrams for the dimension-5 operator $\bar{L} \, \tilde{H} \, \chi \, \nu_{R_a}$, that 
generate Dirac neutrino masses. We have associated each topology to graphs or
Feynman diagrams without taking into account their Lorentz nature. We understand as diagrams 
when the fermion and scalar lines are specified. We consider the case where the neutrino masses
arise after spontaneous breaking of the \SM and $U(1)_{B-L}$ symmetry. Under this requirement the lowest order comes from topologies at dimension-5~\cite{Calle:2018ovc}, as shown in Figure~\ref{fig:gencase}.
This implies that in the diagrams  corresponding to topologies T1 and T2, two external legs are fermions (continuous lines) and the others are scalars (dashed lines). One of these scalars corresponds to the \sm Higgs doublet which is responsible for the EWSB and the other one, the scalar singlet $\chi$, is assigned to break $U(1)_{B-L}$ symmetry.

It turns out that within the T1 topology there is only one possible diagram, called T1-1 in Figure~\ref{fig:gencase}. There are two possible realizations of such a diagram depending on whether the internal fermions are $SU(2)_L$ 
doublets or singlets\footnote{In this work we restrict to only colorless $SU(2)_L$ singlet and doublet representations. Models with intermediate fields carrying non-trivial color, exotic hypercharge or having higher  $SU(2)_L$ representations are also possible but will not be considered here. }.
This is in contrast to the classification provided in~\cite{Calle:2018ovc}.
For the T2 topology three different dimension-5 diagrams (T2-1, T2-2, T2-3) can 
be drawn. In this case the different models that can be constructed also 
depend on the $SU(2)_L$ representations of the internal fields.
It is important to mention that all these diagrams are genuine in the sense that there is
no lower order contribution to neutrino mass. That is, there is no tree-level contribution to the neutrino mass generation out of the matter content for these cases.
Note that within a model where the neutrino mass is generated through a T2-1 diagram 
it can have a T1-1 diagram when the scalar on top is integrated out. Regardless 
this, T2-1 remains being genuine under our previous definition.  Having considered all possible topologies and diagrams for one loop realization of the operator $\bar{L} \, \tilde{H} \, \chi \, \nu_{R_a}$ we now turn to the scenarios where the intermediate particles can belong to a dark sector with the lightest one being a potential dark matter candidate. The dark matter stability in such scenarios again depends crucially on the residual $\mathcal{Z}_n$ symmetry which we discuss with explicit examples in next section.


\section{Residual symmetry and dark matter stability}
\label{sec:symmetryrole}


 We already saw the crucial role the residual symmetries play in determining the Dirac or Majorana nature of neutrinos. They play an equally important role in determining the stability of the potential dark matter candidate in a given model. In this section we elucidate this role further by taking the one loop completion of the dimension-5 operator$\bar{L}_i \, \tilde{H} \, \chi \, \nu_{R_a}$; $i = 1,2,3$ and $a = 1,2$ of previous section and scrutinizing it in more details. 

As discussed in details in Section~\ref{sec:dir-maj}, taking the anomaly free solution with $\nu_R = (-4, -4, 5)$ under $U(1)_{B-L}$ has the advantage that the tree level Yukawa term is automatically forbidden. However, the dimension-5 operator $\bar{L} \, \tilde{H} \, \chi \, \nu_{R_a}$ is allowed by $U(1)_{B-L}$ and paves way for generating small Dirac masses at one loop level as shown in~\cite{Bonilla:2018ynb,Calle:2018ovc}. Once the scalar $\chi$ carrying 3 units of $U(1)_{B-L}$ charge gets vev, the $U(1)_{B-L}$ symmetry is broken to a residual $\mathcal{Z}_n$ subgroup. In such a setup, without adding any extra symmetry, the intermediate particles running in the loop can in principle be arranged to belong to the ``dark sector'' with the lightest of them being potentially stable \textit{\`a la} scotogenic mechanism~\cite{Ma:2006km}.  However, whether such a potential dark matter particle will be stable or not is dictated by the residual $\mathcal{Z}_n$ subgroup and not by the  $U(1)_{B-L}$  symmetry~\cite{Bonilla:2018ynb,CentellesChulia:2019gic}. In this section we demonstrate this fact explicitly by taking  the model realizations of the T1 topology of Section~\ref{sec:one-loop}. A similar analysis can also be done for all other diagrams belonging to the T2 topology. 

The first thing to notice is that when $U(1)_{B-L} \to \mathcal{Z}_n$, the type of residual 
$\mathcal{Z}_n$ subgroup that is left unbroken depends on the vev of the $U(1)_{B-L}$ charge carrying scalar as well as on the details of charges of the intermediate fields required for UV completion.
For the operator $\bar{L}_i \tilde{H} \chi \nu_{R_a}$; $i = 1,2,3$ and $a = 1,2$, since $\chi \sim 3$ under $B-L$, the residual $\mathcal{Z}_n$ subgroup can only be a $\mathcal{Z}_{3m}$; $m\in \mathbb{Z}^+ $ group. However, exactly which  $\mathcal{Z}_{3m}$ is left unbroken after the breaking of the $U(1)_{B-L}$ symmetry,  is dependent on the details of a given model and in particular to the lowest $U(1)_{B-L}$ charge in the model. Now we further illustrate this fact with explicit examples.

To see the crucial role of residual symmetry in dark matter stability,  lets consider the model realizations of the T1-1 diagram of Figure~\ref{fig:gencase}. The diagram can be UV completed by adding  new \SM singlet fermions $N_L, N_R$ and scalar $\xi$ along with the inert $SU(2)_L$ doublet scalar $\eta$. All these fields are charged under $U(1)_{B-L}$ symmetry with their charges as listed in Table~\ref{tab:part-z3}.

\begin{table}[h!]
\begin{center}
\begin{tabular}{| c || c | c | c || c |}
  \hline 
&  \hspace{0.1cm} Fields  \hspace{0.1cm}      & \hspace{0.1cm}   $SU(2)_L \otimes U(1)_Y$   \hspace{0.1cm}         &  \hspace{0.2cm}   $U(1)_{B-L}$  \hspace{0.2cm}   & \hspace{0.5cm} $\mathcal{Z}_3$ \hspace{0.5cm}                             \\
\hline \hline
\multirow{4}{*}{ \begin{turn}{90}\,\,\, Fermions \end{turn} } &
 $L_i$        	      &    ($\mathbf{2}, {-1/2}$)       &   {\color{blue}${-1}$}    &  	{\color{red}${\omega^2}$}                    \\	
&   $\nu_{R_a}$       &   ($\mathbf{1}, {0}$)           & {\color{blue} $-4$ }      &  	{\color{red}${\omega^2}$} \\
&   $\nu_{R_3}$       &   ($\mathbf{1}, {0}$)           & {\color{blue} $5$ }       &  	{\color{red}${\omega^2}$} \\
&   $N_{L(R)}$    	  &   ($\mathbf{1}, {0}$)           & {\color{blue}${q}$ }      &  	{\color{red}${\omega^q}$}   \\
\hline \hline
\multirow{4}{*}{ \begin{turn}{90} Scalars \end{turn} } &
 $H$  		         &  ($\mathbf{2}, {1/2}$)          &  {\color{blue}${0}$ }      &  	{\color{red}${\omega^0}$} \\
& $\chi$          	 &  ($\mathbf{1}, {0}$)            &  {\color{blue}${3}$ }      &  	{\color{red}${\omega^0}$}\\		
& $\eta$          	 &  ($\mathbf{2}, {1/2}$)      &  {\color{blue}${1 - |q|}$}     &  	{\color{red}${\omega^{1 - |q|}}$}      \\
& $\xi$             &  ($\mathbf{1}, {0}$)        &  {\color{blue}${4 - |q|}$}      & 	{\color{red}${\omega^{1 - |q|}}$} \\	
    \hline
  \end{tabular}
\end{center}
\caption{\begin{footnotesize}Matter content and charge assignments for realizing T1-1. For all integer values of $B-L$ charge $q$, the $U(1)_{B-L} \to \mathcal{Z}_3$. The residual $\mathcal{Z}_3$ charges are shown in last column where $\omega = e^{2\pi I/3}; \omega^3 = 1$ is the cube root of unity.                                                                                                                                                                                                                                                                                       \end{footnotesize}}
 \label{tab:part-z3} 
\end{table}

\begin{figure}[H]
    \centering
    \begin{subfigure}[t]{\linewidth}
    \includegraphics[width=\linewidth]{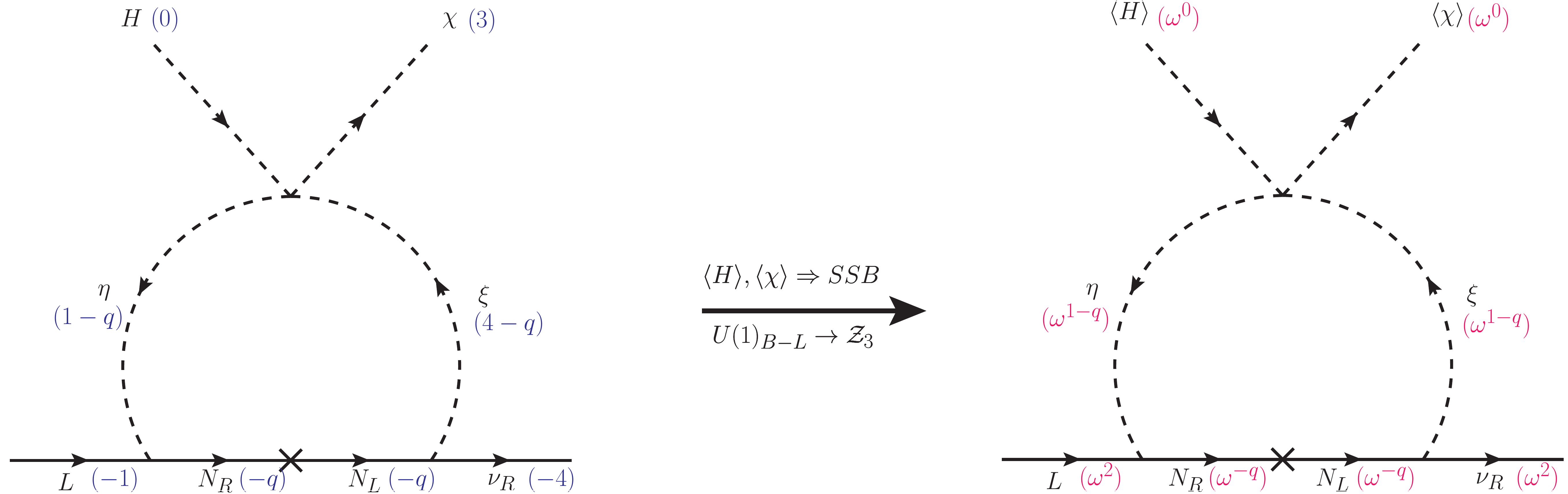}
    \caption{\begin{footnotesize}For the case 
    $q \in \mathbb{Z}^-$ .      \end{footnotesize}}
    \label{fig:gencase-neg-q}
    \end{subfigure}
    \par \vspace{0.5cm}
    \begin{subfigure}[t]{\linewidth}
    \includegraphics[width=\linewidth]{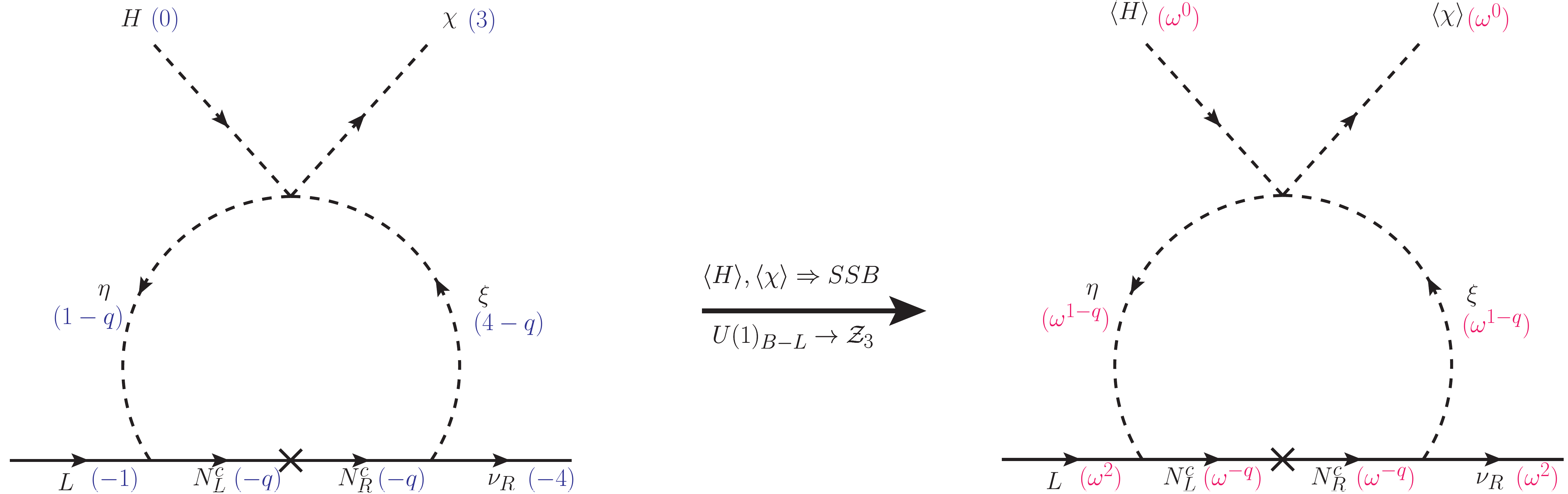}
    \caption{\begin{footnotesize} For the case 
    $q \in \mathbb{Z}^+$ .      \end{footnotesize}}
    \label{fig:gencase-pos-q}
    \end{subfigure}
    \caption{\begin{footnotesize}One loop neutrino mass generation diagrams highlighting the $ U(1)_{B-L} \to \mathcal{Z}_3$ breaking pattern. The $B-L$ charges (left diagram) of the fields  are in blue while the residual $\mathcal{Z}_3$ charges (right diagram) are in red. 
    For  $\mathcal{Z}_3$ symmetry $\omega = e^{2\pi I/3}; \omega^3 = 1$ is the cube root of unity.  
    Notice that the \sm gauge singlet fermions $N_L$ and $N_R$ have switched their roles in the two cases. \end{footnotesize}  }
    \label{fig:gen-q}
\end{figure}

As shown in Table~\ref{tab:part-z3}, the one-loop neutrino mass can be generated for all values of the $U(1)_{B-L}$ charge $q$. However, the residual symmetry in all cases need not be same. To begin with, lets consider the cases where $q \in \mathbb{Z}$ i.e. $q$ only takes integer values. In all such cases the lowest $U(1)_{B-L}$ charge in the model is $\pm 1$, the charge of the lepton doublets $L_i$. Since, $\chi \sim 3$ under $U(1)_{B-L}$, its vev will then break $U(1)_{B-L} \to \mathcal{Z}_3$. The residual $\mathcal{Z}_3$ symmetry, being an odd $\mathcal{Z}_n$ group, is enough to protect the Dirac nature of neutrinos in accordance with \eqref{eq:oddzn}. The one loop Dirac neutrino mass realization for $q \in \mathbb{Z}^-$ and $q \in \mathbb{Z}^+$ are shown in Figure~\ref{fig:gencase-neg-q} and in Figure~\ref{fig:gencase-pos-q}, respectively. Notice that the \SM singlet fermions $N_L$ and $N_R$ simple switch roles for the $q \in \mathbb{Z}^-$ and $q \in \mathbb{Z}^+$ cases.

Notice that the neutral components of the intermediate particles in the loops of Figure~\ref{fig:gen-q} can in principle belong to the ``dark sector'' with the lightest of them being a good dark matter candidate
\textit{\`a la} scotogenic model. Although the residual $\mathcal{Z}_3$ symmetry is enough to protect the Dirac nature of neutrinos, it is not enough to protect the stability of a potential dark matter candidate. In fact, as has been argued in~\cite{Bonilla:2018ynb}, any odd $\mathcal{Z}_n$ subgroup of $U(1)_{B-L}$, on its own cannot protect the dark matter stability in models employing scotogenic mechanism. The dark matter stability in all such models has to be achieved either by adding another explicit symmetry or by choosing ``exotic''  $U(1)_{B-L}$ charges for the intermediate fields such that an accidental symmetry also appears in the model. We now further elaborate on this by looking case by case at models with $q = 0, \cdots \pm 6$. We start with looking at the cases where the dark matter stability is  completely lost and then move on to the cases where a new accidental symmetry comes into play.


\subsubsection{$q = \pm 1$ case}

\begin{figure}[!htbp]
\begin{center}
 \begin{tabular}{lr}
\hspace{-1.5cm}
\\ \begin{subfigure}[b]{1\textwidth}
\includegraphics[width=0.5\linewidth]{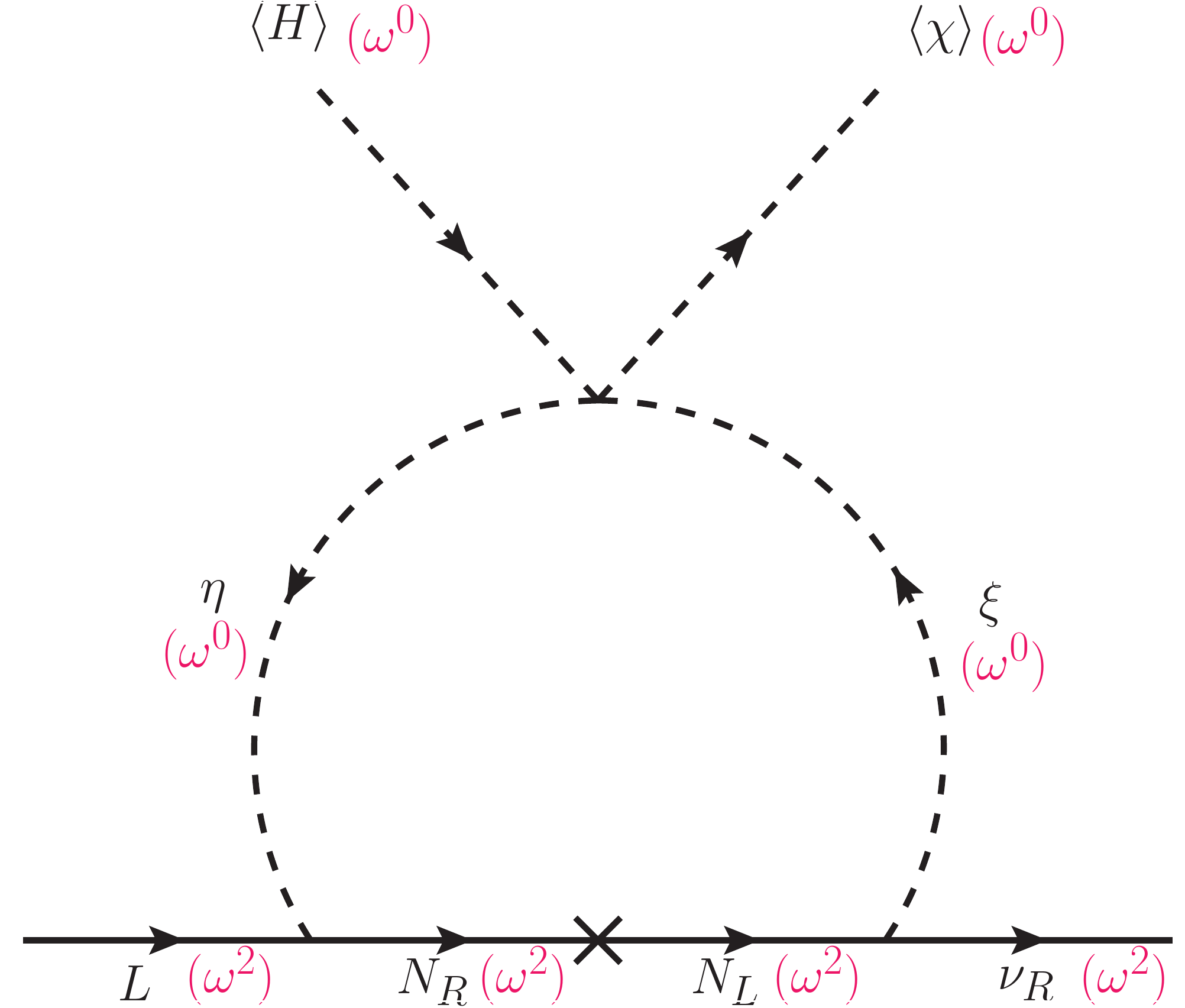}
\caption[]{\footnotesize The $q = - 1$ case, raditative neutrino mass generation diagram along with residual $\mathcal{Z}_3$ charges. Analogous diagram can be drawn for  $q = + 1$ case, following Figure~\ref{fig:gencase-pos-q}.    }
         \label{fig:q1}
         \end{subfigure}
\quad \quad \qquad
\\
\begin{subfigure}[b]{1\textwidth}
\includegraphics[width=1\linewidth]{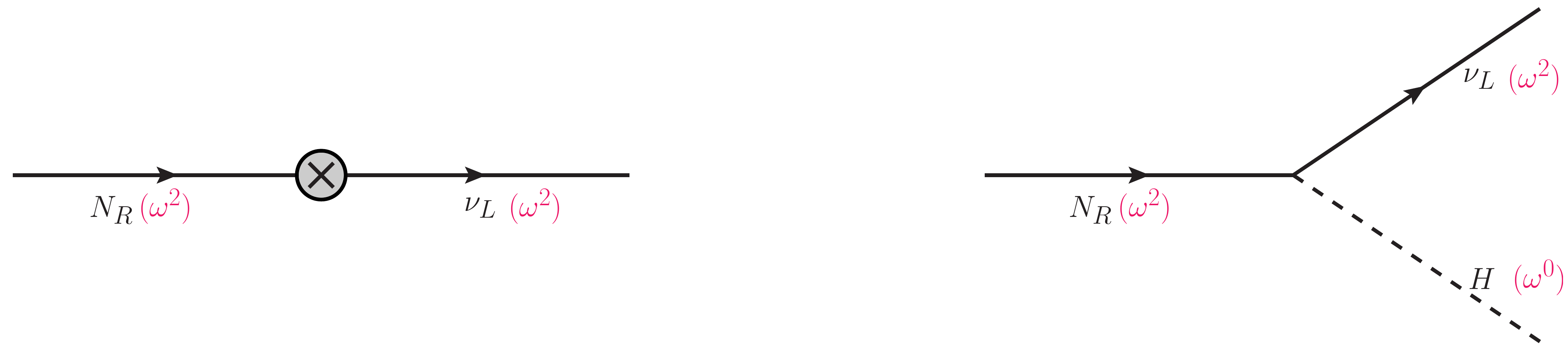}
\caption[a]{\footnotesize The mixing and decay diagrams of dark sector particles, ultimately leading to the decay of potential dark matter candidate.}
         \label{fig:dmq1}
\end{subfigure} \\ 
 \end{tabular}
 \end{center}
\vspace{-0.6cm}        
\caption{ \footnotesize The raditative neutrino mass generation diagram and the mixing and decay diagrams of dark sector particle, for the $q = - 1$ case. Anologous diagrams can also be drawn for $q = + 1$ case with $N_L$ and $N_R$ switching their roles like in Figure~\ref{fig:gencase-neg-q} and Figure~\ref{fig:gencase-pos-q} }
\label{fig:q1-case}
\end{figure}

Let's start with the simplest case of $q = \pm 1$. For definiteness we take the case of $q = -1$ but analogous discussion can be carried out for $q = +1$ case as well.  The one loop neutrino mass generation diagram along with the residual $\mathcal{Z}_3$ charges is shown in Figure~\ref{fig:q1}.

However, for this case the Yukawa term $\bar{L} \tilde{H} N_R$ is also allowed by the symmetry. This term leads to mixing as well as decay of the $N_R$ fields as shown in Figure~\ref{fig:dmq1}.  Since $N_R$
is a ``dark sector'' field, this coupling ultimately provides a channel for the dark matter decay, even if $N_R$ itself is the not the dark matter candidate. This is because just like in scotogenic model, here also, all intermediate fields can ``decay'' among each other.  If one of them has a decay channel to \sm particles or to the new external fields $\nu_R, \chi$, then any potential dark matter candidate will ultimately decay by first going to real/virtual $N_R$ which will finally decay to $\nu_L$ and $H$ as shown in Figure~\ref{fig:dmq1}.


\subsubsection{$q = \pm 2$ case}

\begin{figure}[!htbp]
\begin{center}
 \begin{tabular}{lr}
\hspace{-1.5cm}
\\ \begin{subfigure}[b]{0.45\textwidth}
\includegraphics[width=1\linewidth]{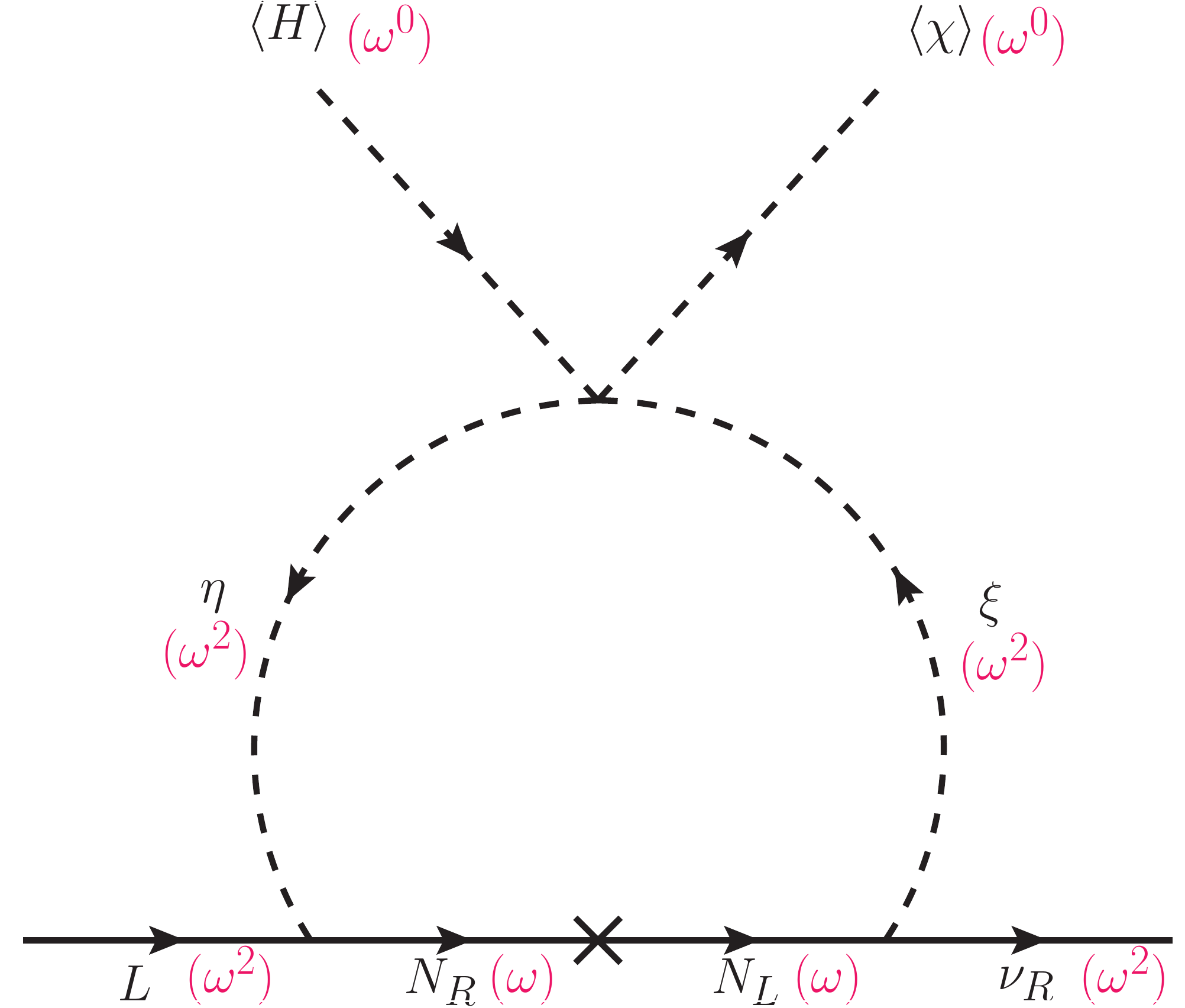}
\caption[]{Neutrino mass diagram for $q = - 2$ case. }
         \label{fig:q2}
         \end{subfigure}
\quad \quad \qquad
\begin{subfigure}[b]{0.45\textwidth}
\includegraphics[width=1\linewidth]{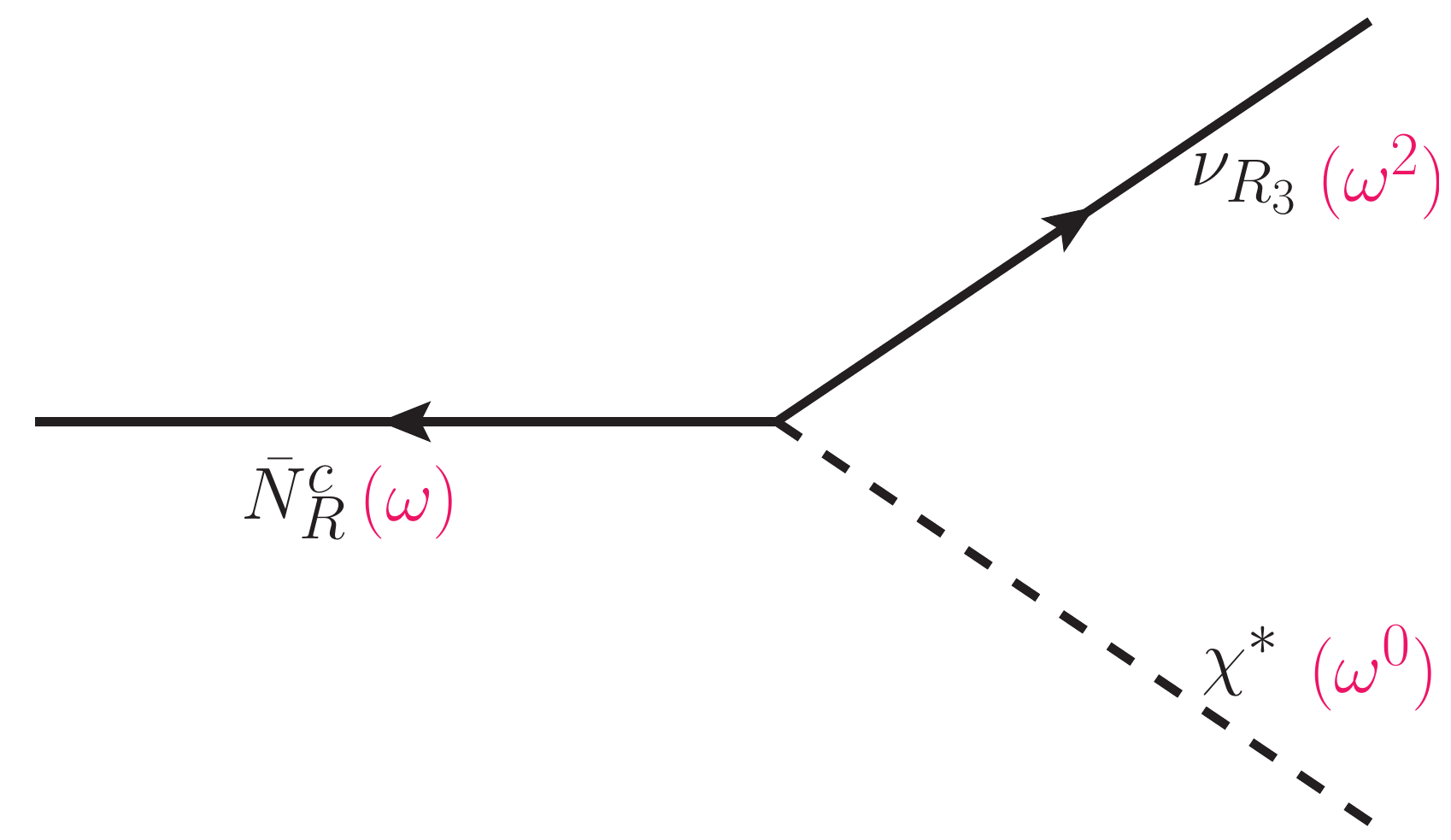}
\caption[a]{The decay diagram of dark sector particle.}
         \label{fig:dmq2}
\end{subfigure} \\ 
 \end{tabular}
 \end{center}
\vspace{-0.6cm}        
\caption{ \footnotesize The $q = - 2$ case, raditative neutrino mass generation diagram along with residual $\mathcal{Z}_3$ charges.  Analogous diagram can be drawn for  $q = + 2$ case, following Figure~\ref{fig:gencase-pos-q}.  Shown also, the decay diagram of dark sector particle, ultimately leading to the decay of potential dark matter candidate. }
\label{fig:q2-case}
\end{figure}

For the $q = - 2$ case, the Feynman diagram leading to neutrino mass generation along with the residual $\mathcal{Z}_3$ charges of the fields is shown in Figure~\ref{fig:q2}.  As before the case of $q=2$ can be analogously discussed with $N_L$ and $N_R$ fields switching their roles.

In this case one can again write down a dimension-4 term $\bar{N}^c_R \nu_{R_3} \chi^*$ which leads to the decay of $N_R$ field as shown in Figure~\ref{fig:dmq2}. As can be seen from Figure~\ref{fig:dmq2}, the residual $\mathcal{Z}_3$ symmetry does not forbid this decay. Now, since $\bar{N}^c_R$ is a dark sector particle, its decay channel ultimately provides a decay path for any potential dark matter  candidate. Thus, the dark matter stability is again lost.


\subsubsection{$q = \pm 3$ case}

\begin{figure}[!htbp]
\begin{center}
 \begin{tabular}{lr}
\hspace{-1.5cm}
\\ \begin{subfigure}[b]{0.45\textwidth}
\includegraphics[width=1\linewidth]{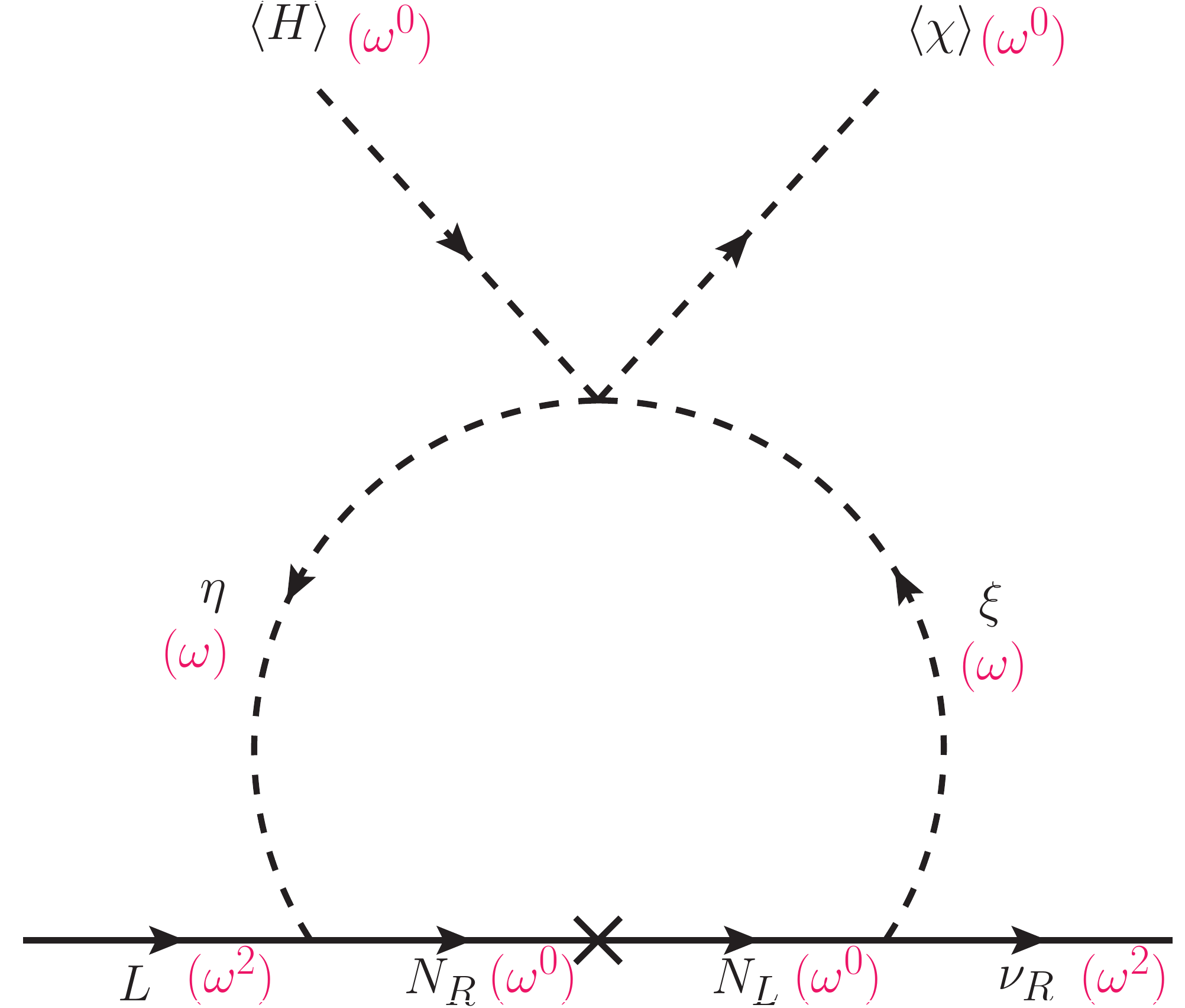}
\caption[]{Neutrino mass diagram for $q = - 3$ case. }
         \label{fig:q3}
         \end{subfigure}
\quad \quad \qquad
\begin{subfigure}[b]{0.45\textwidth}
\includegraphics[width=1\linewidth]{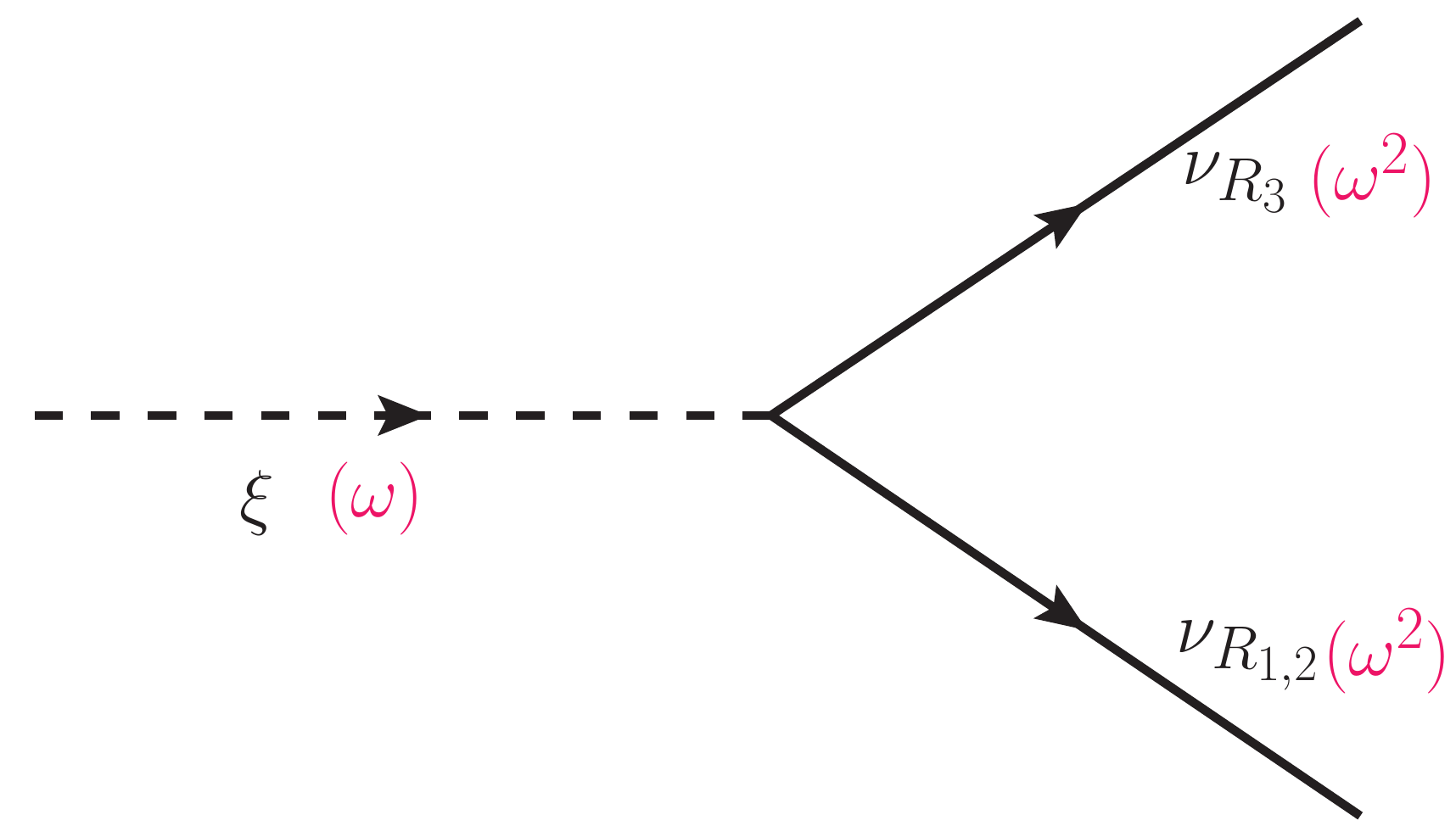}
\caption[a]{The decay diagram of dark sector particle.}
         \label{fig:dmq3}
\end{subfigure} \\ 
 \end{tabular}
 \end{center}
\vspace{-0.6cm}        
\caption{ \footnotesize The $q = - 3$ case for raditative neutrino mass generation and decay channel of dark sector particles. Analogous diagram can be drawn for  $q = + 3$ case. }
\label{fig:q3-case}
\end{figure}

Turning to the $q = \pm 3$, the Feynman diagram for neutrino mass generation is shown in Figure~\ref{fig:q3}. The diagram in Figure~\ref{fig:q3} is drawn for the case $q = -3$ but a similar  diagram can also be drawn for $q = +3$ case.

In this case the  dimension-4 operator $\xi \nu_{R_{a}}\nu_{R_3} $ connecting intermediate scalar $\xi$ to right handed neutrinos $\nu_{R_a} \nu_{R_3}$ is allowed by the symmetries.  This operator leads to decay of $\xi$ to right handed neutrinos as shown in Figure~\ref{fig:dmq3}. As can be seen from 
Figure~\ref{fig:dmq3}, this decay is again allowed by the residual $\mathcal{Z}_3$ symmetry.
Since $\xi$ is a dark sector field, its decay ultimately implies decay of any potential dark matter candidate. Thus, the dark matter stability is again explicitly lost.


\subsubsection{$q = \pm 4$ case}

\begin{figure}[!htbp]
\begin{center}
 \begin{tabular}{lr}
\hspace{-1.5cm}
\\ \begin{subfigure}[b]{0.45\textwidth}
\includegraphics[width=1\linewidth]{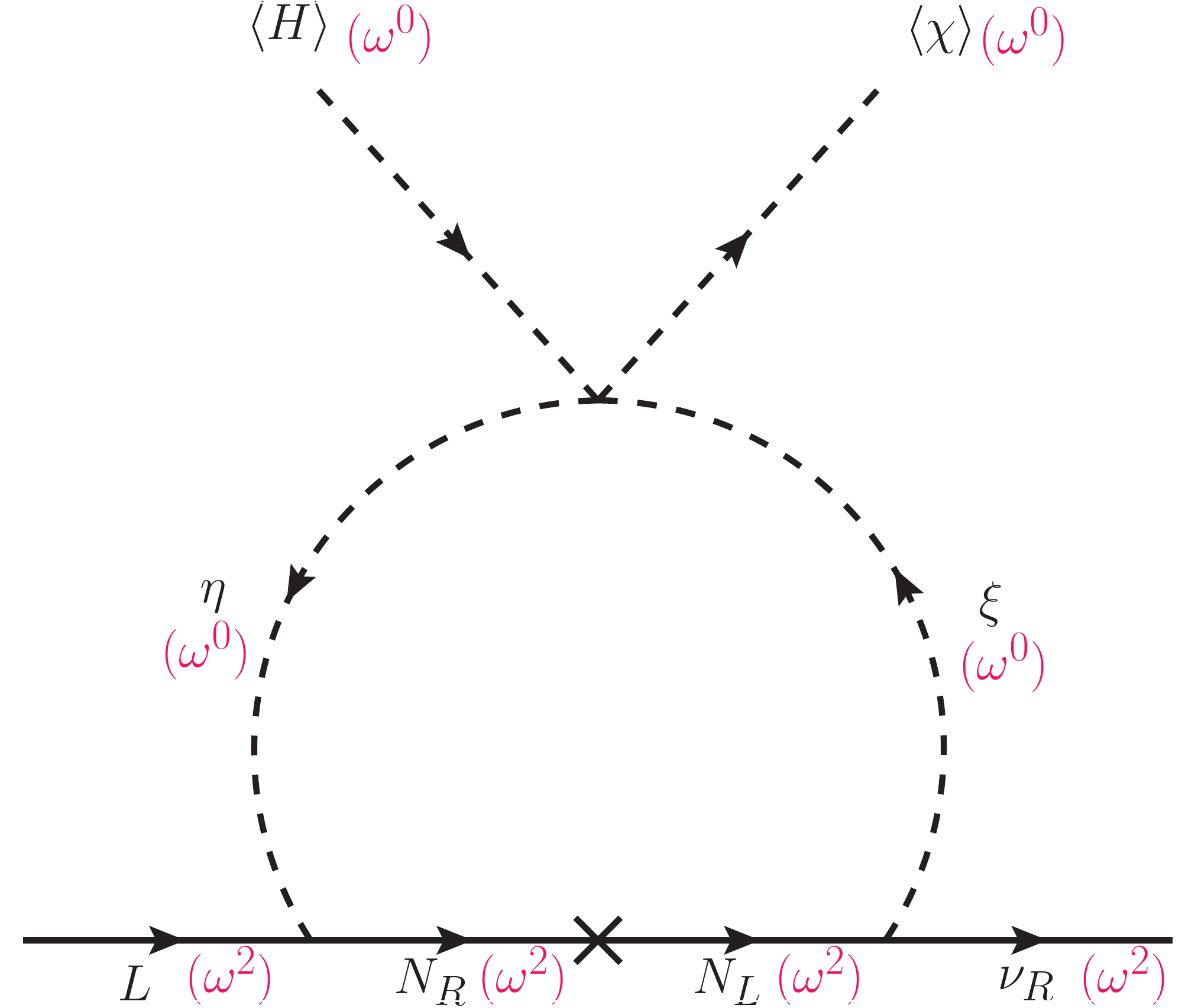}
\caption[]{Neutrino mass diagram for $q = - 4$ case. }
         \label{fig:q4}
         \end{subfigure}
\quad \quad \qquad
\begin{subfigure}[b]{0.45\textwidth}
\includegraphics[width=1\linewidth]{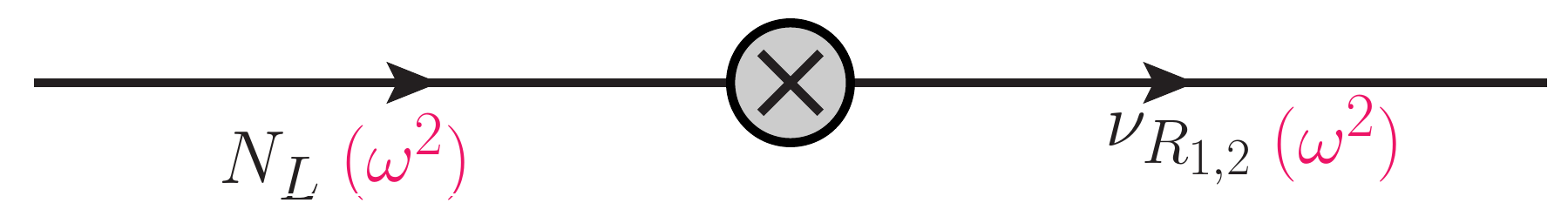}
\caption[a]{The mixing diagram of dark sector particle.}
         \label{fig:dmq4}
\end{subfigure} \\ 
 \end{tabular}
 \end{center}
\vspace{-0.6cm}        
\caption{ \footnotesize The $q = - 4$ case for raditative neutrino mass generation and mixing diagram of dark sector particle. Analogous diagram can be drawn for  $q = + 4$ case. }
\label{fig:q4-case}
\end{figure}

For the case of  $q = - 4$, the neutrino mass diagram is as shown in Figure~\ref{fig:q4}. A similar diagram can also be drawn for the $q = +4$ case as well.
    
In this case there is a direct coupling $\bar{N}_L \nu_{R_{a}}$ between the intermediate field $N_L$ and the right handed neutrinos $\nu_{R_{a}}$ as shown in Figure~\ref{fig:dmq4}. This coupling leads to mixing between the two fields which is allowed by the residual $\mathcal{Z}_3$ symmetry. Owing to this mixing, 
all the dark sector particles, including any potential dark matter candidate, again have a decay channel available to them. Thus, in this case as well, the dark matter stability is explicitly lost.
\newpage


\subsubsection{$q = \pm 5$ case}

\begin{figure}[!htbp]
\begin{center}
 \begin{tabular}{lr}
\hspace{-1.5cm}
\\ \begin{subfigure}[b]{0.45\textwidth}
\includegraphics[width=1\linewidth]{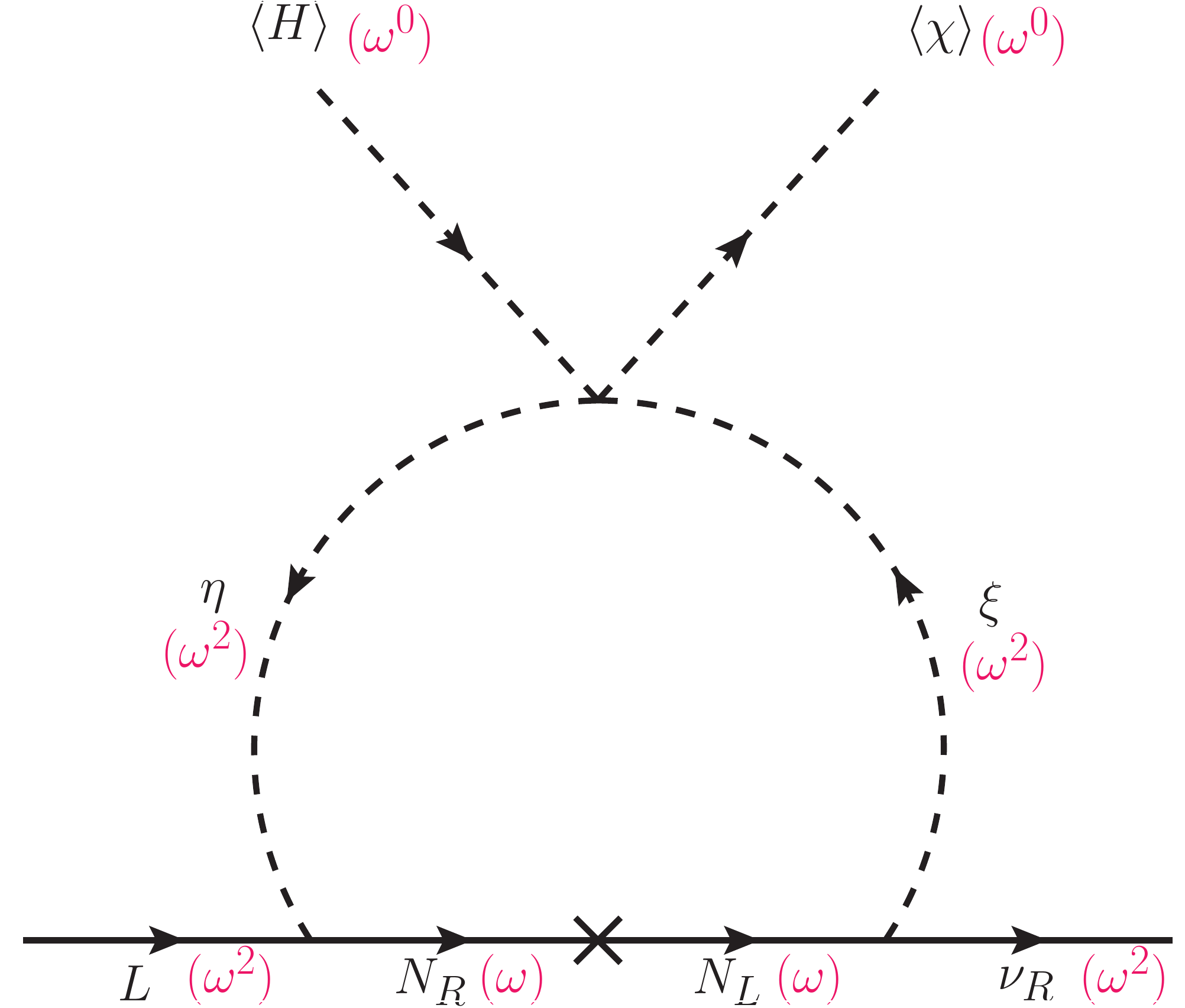}
\caption[]{Neutrino mass diagram for $q = - 4$ case. }
         \label{fig:q5}
         \end{subfigure}
\quad \quad \qquad
\begin{subfigure}[b]{0.45\textwidth}
\includegraphics[width=1\linewidth]{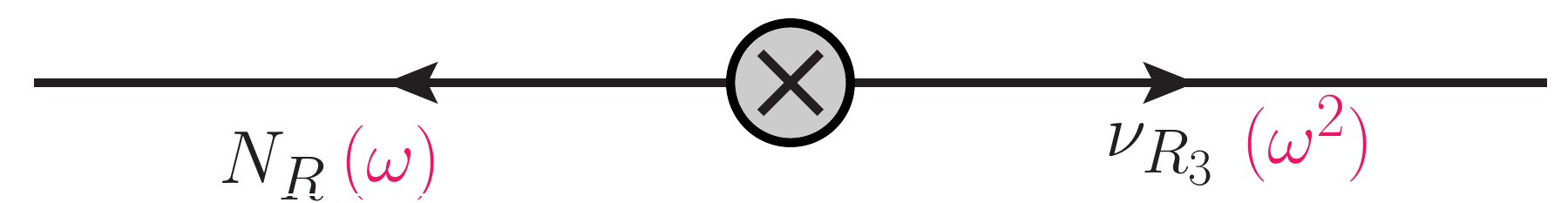}
\caption[a]{The mixing diagram of dark sector particle.}
         \label{fig:dmq5}
\end{subfigure} \\ 
 \end{tabular}
 \end{center}
\vspace{-0.6cm}        
\caption{ \footnotesize The $q = - 5$ case for raditative neutrino mass generation and mixing diagram of dark sector particle. Analogous diagram can be drawn for  $q = + 5$ case. }
\label{fig:q5-case}
\end{figure}

The Feynman diagram leading to the neutrino mass generation for $q = - 5$ is shown in Figure~\ref{fig:q5}.
A very similar diagram can also be drawn for the $q = + 5$ case with the fields $N_L$ and $N_R$ switching their roles.

In this case, analogous to the $q = \pm 4$ case, their is a direct coupling $\bar{N}^c_R \nu_{R_{3}}$
allowed by the symmetries. This leads to the mixing between these two fields as shown in Figure~\ref{fig:dmq5}. Again this mixing is allowed by the residual $\mathcal{Z}_3$ symmetry and it ultimately leads to the decay of any potential dark matter candidate.

\subsubsection{Accidental symmetry: The $q = 0$ and $q = \pm 6$ cases}

Finally we turn to the cases where an additional accidental symmetry of the model leads to dark matter stability. The presence of accidental symmetries as well as stability of particles owing to accidental symmetries is not a new concept. In the \sm the Lepton $U(1)_L$ and Baryon number $U(1)_B$ appear as accidental symmetries of the model~\cite{Weinberg:1980bf}. As we already saw, the combination $U(1)_{B-L}$ and its residual symmetries are intimately connected with the Dirac/Majorana nature of neutrinos. The other combination $U(1)_{B+L}$ is responsible for stability of proton. Its breaking pattern and residual symmetries are intimately connected with not only stability of proton but also dictate the possible proton decay channels~\cite{Nath:2006ut,Fonseca:2018ehk,Reig:2018yfd}. Just like the accidental $U(1)_{B+L}$ protects proton decay in \sm, in many dark matter models an accidental symmetry can protect dark matter decay~\cite{Cirelli:2005uq,Wang:2017mcy,Calle:2018ovc}. A well know example of this is the case of ``minimal dark matter'' where even after the $\mathrm{SU(2)_L \otimes U(1)_Y} \to \mathrm{U(1)_{em}}$, dark matter remains stable thanks to an accidental symmetry present in the model~\cite{Cirelli:2005uq}.  

\begin{figure}[!htbp]
\begin{center}
 \begin{tabular}{lr}
\hspace{-1.5cm}
\\ \begin{subfigure}[b]{0.45\textwidth}
\includegraphics[width=1\linewidth]{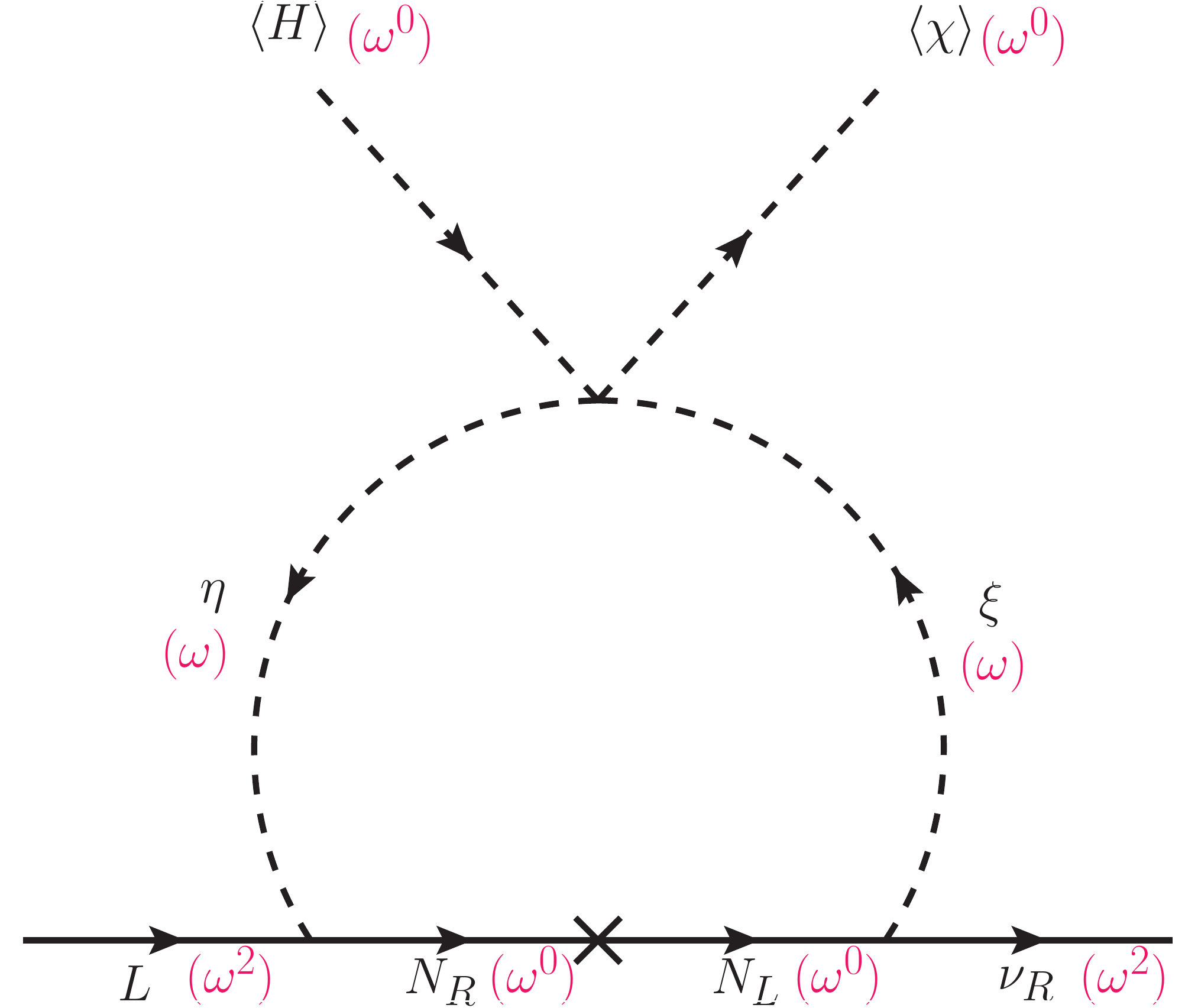}
\caption[]{Neutrino mass diagram for $q = 0$ case. }
         \label{fig:q0}
         \end{subfigure}
\quad \quad \qquad 
\begin{subfigure}[b]{0.45\textwidth}
\includegraphics[width=1\linewidth]{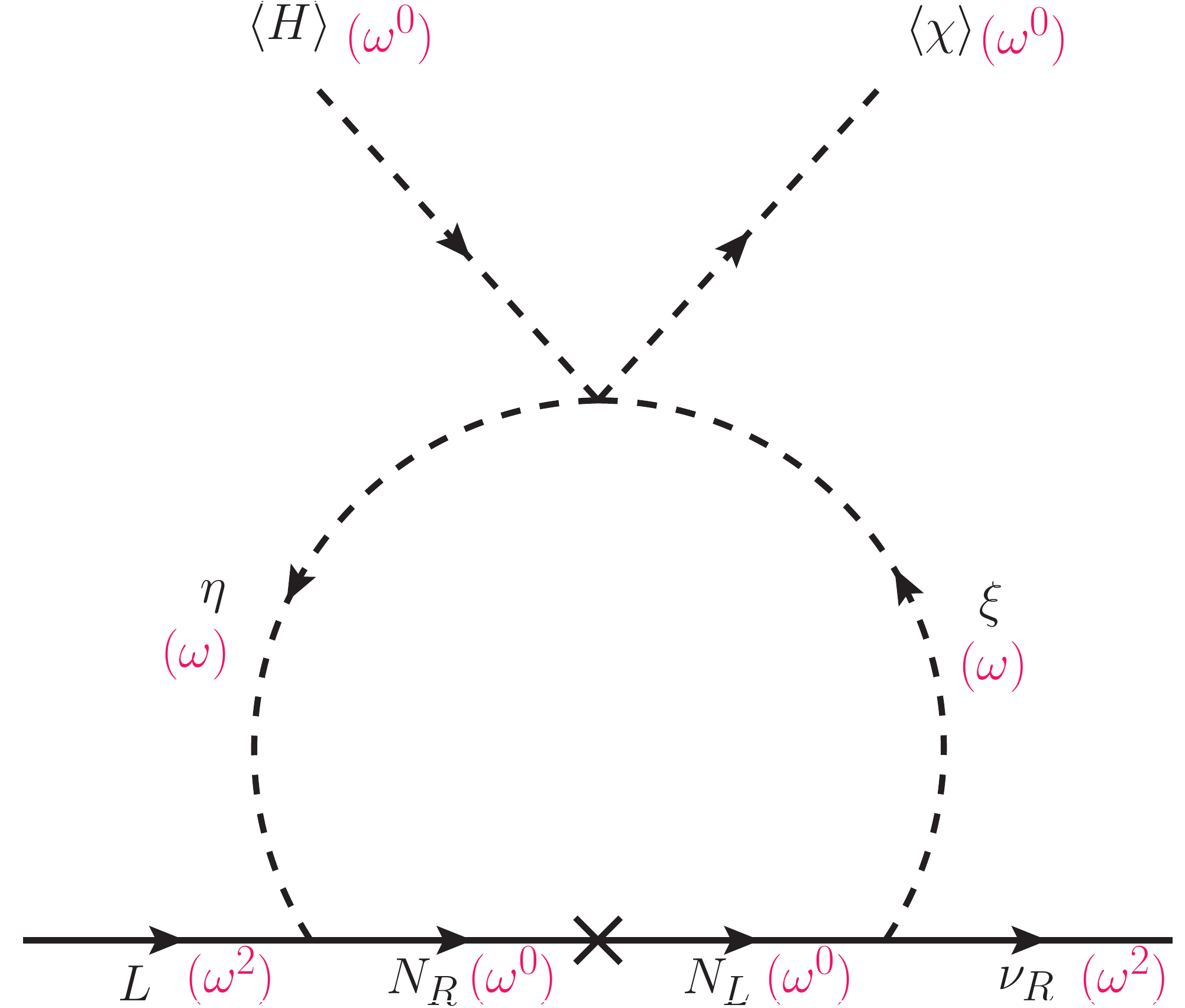}
\caption[a]{Neutrino mass diagram for $q = -6$ case.}
         \label{fig:q6}
\end{subfigure} \\ 
 \end{tabular}
 \end{center}
\vspace{-0.6cm}        
\caption{ \footnotesize The neutrino mass generation diagrams for both $q = 0$ and $q = -6$ cases. In both these cases the internal fermions do not carry non-trivial $\mathcal{Z}_3$ charges. However, one can still have a stable dark matter thanks to the presence of an extra accidental symmetry in the models.
 }
\label{fig:q0-q6-case}
\end{figure}

In our setup where $U(1)_{B-L} \to \mathcal{Z}_3$, the cases of $q=0$ and $q = \pm 6$ provide such examples. The  Feynman diagrams for neutrino mass generation are shown in Figure~\ref{fig:q0-q6-case}. As can be seen from Figure~\ref{fig:q0-q6-case}, the intermediate fermions $N_L, N_R$ do not carry any charge under the residual $\mathcal{Z}_3$ symmetry and hence their decays are not protected by it. Furthermore, although the scalar field $\xi$ does carry $\mathcal{Z}_3$ charge, still its decay $\xi \to \nu_R$ is allowed by the residual symmetry. Thus, the dark matter stability in both these cases is not protected by the residual $\mathcal{Z}_3$ symmetry. Still the dark matter particle in both these cases is stable thanks to presence of an accidental symmetry in the model.

The presence of an accidental symmetry in both these cases can be seen from the fact that in the unbroken phase the $U(1)_{B-L}$ symmetry allows one to write down decay terms like $\eta^\dagger H \nu_{R_a} \nu_{R_3}$ (for $q = 0$) and $\xi \chi \nu^c_{R_a} \nu^c_{R_3}$ (for $q = \pm 6$). Yet with the particle content of these models, see Table~\ref{tab:part-z3}, these decay operators cannot be UV completed, clearly indicating presence of an accidental symmetry stabilizing dark matter. This situation is akin to the proton stability in \sm where again the gauge symmetries allow non-renormalizable proton decay operators which cannot be UV completed with the particle content of \sm~\cite{Weinberg:1980bf}.  In fact, similar analysis can be carried out for $q < -6$ or $q > 6$. In each case it can be shown that the residual $\mathcal{Z}_3$ alone cannot stabilize dark matter. In fact it has been shown in~\cite{Bonilla:2018ynb}, arguing at very general grounds that any odd residual $\mathcal{Z}_n$ symmetry and in particular 
$\mathcal{Z}_3$ symmetry cannot stabilize dark matter. However, appropriately chosen even $\mathcal{Z}_n$ residual symmetries can indeed stabilize dark matter while simultaneously protecting the Dirac nature of neutrinos. In next section we present one such example. For the most general treatment we refer to~\cite{Bonilla:2018ynb}.   
%

\section{Dirac neutrinos and dark matter Stability from a residual symmetry: $U(1)_{B-L} \to \mathcal{Z}_6$}
\label{sec:model}

We already saw that $U(1)_{B-L} \to \mathcal{Z}_3$ can protect the Dirac nature of neutrinos but $\mathcal{Z}_3$ being an odd group, cannot provide dark matter stability. However, if  $U(1)_{B-L} \to \mathcal{Z}_{2m}$, $m \in \mathbb{Z}^+, m \geq 3$; then one can indeed have dark matter stability as well as protect Dirac nature of neutrinos in a scotogenic like setup~\cite{Bonilla:2018ynb}. In this section we show this by explicitly constructing one such example where:
\begin{enumerate}
\item[ I.] Neutrinos are Dirac in nature.
\item[ II.]  Neutrino mass is generated at one loop level. 
\item[ III.] The intermediate particles in the loop belong to a ``dark sector'' with the lightest particle among them being a good candidate for stable dark matter. 
\end{enumerate}
We achieve this by  $U(1)_{B-L} \to \mathcal{Z}_6$ in the $\nu_R = (-4,-4,5)$ setup with the residual $\mathcal{Z}_6$ protecting both Dirac nature of neutrinos and dark matter stability. An alternative model to the one considered here, was also constructed in~\cite{Bonilla:2018ynb} which the interested reader can consult for more details\footnote{For anologous treatment in case of Majorana neutrinos see~\cite{CentellesChulia:2019gic}.}.

To start with, recall that as discussed before, since the scalar field $\chi \sim 3$ under $U(1)_{B-L}$ symmetry, its vev breaks $U(1)_{B-L} \to \mathcal{Z}_{3m}$; $m \in \mathbb{Z}^+$. However, exactly which residual $ \mathcal{Z}_{3m}$ symmetry is left unbroken depends on the details of the UV completion and in particular on what is the lowest $U(1)_{B-L}$ charge in the full UV complete theory.
In the case when the \sm lepton doublets $L_i$ carry the lowest charge i.e. the lowest charge is $\pm 1$, then the residual symmetry is $ \mathcal{Z}_{3}$. But in case when there is a smaller charge present in the UV complete model, the residual symmetry will be different. For example, if the lowest $U(1)_{B-L}$ charge in the model is $\pm 1/2$ then the residual symmetry becomes $ \mathcal{Z}_{6}$
while for the lowest charge of $\pm 1/3$, the residual symmetry will be $ \mathcal{Z}_{9}$, and so on.
We will exploit this feature now to break $U(1)_{B-L} \to \mathcal{Z}_6$ by introducing intermediate particles with $\pm 1/2$ charges.

For sake of uniformity and to allow quick comparison, we again take the first topology described in Section~\ref{sec:one-loop} and construct a topologically similar model to the ones considered in Section~\ref{sec:symmetryrole}. 
We construct a \sm extension where the residual symmetry $\mathcal{Z}_6$, stemming from
the spontaneous symmetry breaking of $U(1)_{B-L}$, protects the stability
of the dark matter candidate and the Dirac nature of neutrinos. As before we
demand that the model is anomaly free. Hence, again the simplest option
is to consider that the RH-neutrinos transform as $(\nu_{R_a},\nu_{R_3})\sim(-4,5)$ (with $a=1,2$) under the $U(1)_{B-L}$ symmetry. In order to complete the diagram, as before,
we need to introduce three additional scalars, $\chi$, $\xi$ and $\eta$. The former one 
is responsible for breaking the $U(1)_{B-L}$ symmetry and the other two will be required to UV complete the model and will belong to the dark sector.
A pair of neutral fermions $N_L, N_R$ are also needed and these along with the scalars $\xi$ and 
$\eta$ are part of the dark sector with the lightest of them being a good dark matter candidate.
In Table~\ref{modelZ6} we summarize the matter content and the charge assignments 
of the model. The right column with $\mathcal{Z}_{6}$ label indicates the charge of the 
particles under the residual $\mathcal{Z}_6$ symmetry.

\begin{table}[H]
\begin{center}
\begin{tabular}{| c || c | c | c || c |}
  \hline 
&  \hspace{0.1cm} Fields  \hspace{0.1cm}      & \hspace{0.1cm}   $SU(2)_L \otimes U(1)_Y$   \hspace{0.1cm}         &  \hspace{0.2cm}   $U(1)_{B-L}$  \hspace{0.2cm}   & \hspace{0.5cm} $\mathcal{Z}_6$ \hspace{0.5cm}                             \\
\hline \hline
\multirow{4}{*}{ \begin{turn}{90} Fermions \end{turn} } &
 $L_i$        	  &    ($\mathbf{2}, {-1/2}$)       &   {\color{blue}${1}$}    	  &	 {\color{red}$\omega^4$}                     \\	
&   $\nu_{R_a}$       &   ($\mathbf{1}, {0}$)      & {\color{blue} $-4$ }   &  	 {\color{red}$\omega^4$}\\
&   $\nu_{R_3}$    	  &   ($\mathbf{1}, {0}$)      & {\color{blue}${5}$ }   &    {\color{red} $\omega^4$}     \\
&  $N_{L(R)}$     	  &  ($\mathbf{1}, {0}$) 	     & {\color{blue} ${1/2}$ } &  {\color{red}$\omega$}     \\
\hline \hline
\multirow{4}{*}{ \begin{turn}{90} Scalars \end{turn} } &
 $H$  		 &  ($\mathbf{2}, {1/2}$)      &  {\color{blue}${0}$ }    & {\color{red} $1$}    \\
& $\chi$          	 &  ($\mathbf{1}, {0}$)        &  {\color{blue}${3}$ }  &  {\color{red} $1$}     \\		
& $\eta$          	 &  ($\mathbf{2}, {1/2}$)      &  {\color{blue}${3/2}$}    &  {\color{red}$\omega^3$}       \\
& $\xi$             &  ($\mathbf{1}, {0}$)        &  {\color{blue}${9/2}$}      &	{\color{red}$\omega^3$} \\	
    \hline
  \end{tabular}
\end{center}
\caption{Matter content and charge assignments of the model. Here $\mathcal{Z}_6$ is the residual symmetry with $\omega = e^{2\pi I/6}; \omega^6=1$.}
 \label{modelZ6} 
\end{table}

Notice that the intermediate particles carry $U(1)_{B-L}$ charges in units of $1/2$. Owing to presence of these particles, the vev of $\chi$ now breaks $U(1)_{B-L} \to \mathcal{Z}_6$. The charges of the fields under the residual $\mathcal{Z}_6$ are also listed in Table~\ref{modelZ6}. Since the lepton doublets $L_i$ transform as $\omega^4$ under $\mathcal{Z}_6$, therefore in accordance with \eqref{eq:oddzn} and \eqref{evenzndir}, the neutrinos will be Dirac fermions with their Dirac nature protected by the $\mathcal{Z}_6$ symmetry.

The invariant Lagrangian of the theory that describes 
the neutrino interactions is given by, 
\begin{equation}
{\cal L}_\nu=y^\nu \bar{L}\tilde{\eta}N_{R}+y^{\nu\prime}\bar{N}_{L}\nu_{R}\xi+M \bar{N}_{R}N_{L}+h.c.,
\label{eq:YukInt}
\end{equation}
where $\tilde{\eta} = i \tau_2 \eta^*$ and we are omitting flavour indices for convenience. 
The relevant term, in the scalar potential, to generate the Dirac neutrino mass is given by
\begin{equation}
\label{Veps}
{\cal V}\supset \lambda_D H^\dagger \eta \chi \xi^* + h.c.,
\end{equation}
where $\lambda_D$ is an dimensionless  quartic coupling. 

\begin{figure}[H]
    \centering
    \includegraphics[width=\linewidth]{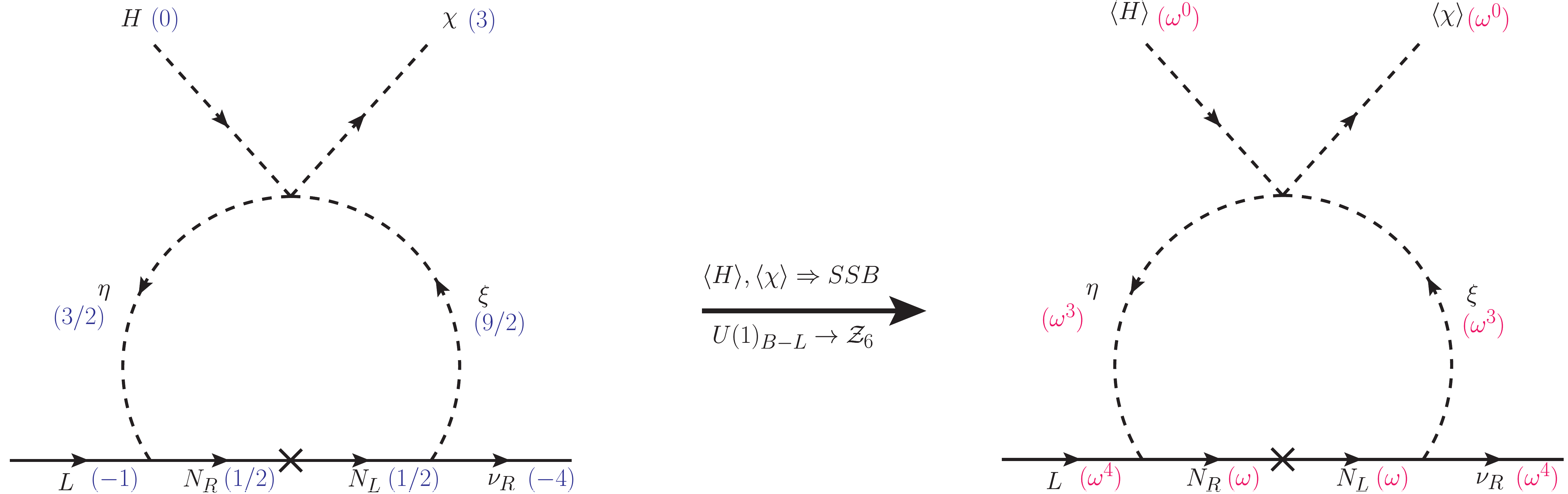}
    \caption{\begin{footnotesize}One loop neutrino mass generation diagram highlighting the $ U(1)_{B-L} \to \mathcal{Z}_6$ breaking pattern. The $B-L$ charges (left diagram) of the fields are given in blue while the residual $\mathcal{Z}_6$ charges (right diagram) are in red. For  $\mathcal{Z}_6$ symmetry $\omega = e^{2\pi I/6}; \omega^6 = 1$ is the six-th root of unity.                                                                       \end{footnotesize}}
    \label{fig:z6}
\end{figure}

After spontaneous symmetry breaking two neutrinos acquire a mass through the loop depicted 
in Figure~\ref{fig:z6}. 

Note that only two RH-neutrinos, $\nu_{R_1}$ and $\nu_{R_2}$,
get mass after spontaneous symmetry breaking. The third one $\nu_{R_3}$ remains massless which is consistent with the current experimental data~\cite{deSalas:2017kay}. Furthermore, it is trivial to extend this simple model by adding another vev carrying scalar $\chi_6 \sim 6$ under  $ U(1)_{B-L}$ to generate mass for the third neutrino.  As mentioned before, since the $ U(1)_{B-L}$ charge of $\chi_6$
is just an integer multiple of the charge of $\chi$ field, presence of $\chi_6$ in the model will not change the $ U(1)_{B-L} \to \mathcal{Z}_6$ breaking pattern.  

Apart from protecting the Dirac nature of neutrinos, the residual $\mathcal{Z}_6$ symmetry also protects the dark matter from decay. To see this, notice that under the $\mathcal{Z}_6$ symmetry all intermediate particles carry odd units of the fundamental charge $\omega$, while all the \sm particles as well as $\nu_R$ and $\chi$ carry even units of  $\omega$ charge. Thus, just like in scotogenic models, here also the intermediate particles split into a separate ``dark sector'' with the decay of any dark sector particle to only \sm particles or to $\nu_R$ and $\chi$, forbidden by the residual  
$\mathcal{Z}_6$ symmetry. Therefore, the lightest of the dark sector particles becomes a good candidate for dark matter with its stability protected by the residual $\mathcal{Z}_6$ symmetry. An alternative model employing similar mechanism is also discussed in~\cite{Bonilla:2018ynb} which can be consulted for more details. For a similar treatment for Majorana neutrinos, we refer to~\cite{CentellesChulia:2019gic}.

\section{Conclusions}

We have discussed the importance of residual symmetries for models that try to make a connection between dark matter stability and the Dirac nature of neutrinos. We showed that in absence of any other conserved symmetry beyond \sm symmetries, the Dirac/Majorana nature of neutrinos is dictated by the residual $\mathcal{Z}_n$ symmetry left unbroken after $U(1)_{B-L}$ breaking. We then discussed the possible one loop realization of the models that can be constructed where Dirac neutrino mass has a dynamical origin at the 1-loop level. These models employ the chiral solutions to $U(1)_{B-L}$ anomaly free solution. Therefore, the tree level dimension-4 Yukawa term between left and right neutrinos is automatically forbidden by the $U(1)_{B-L}$. The small Dirac mass for neutrinos is then generated through one loop realizations of the dimension-5 operator $\bar{L} \tilde{H} \chi \nu_R$.

For such one loop models we also discussed how the same residual $\mathcal{Z}_n$ symmetry protecting the Dirac nature of neutrinos, can also protect the dark matter stability by employing a scotogenic like mechanism. We showed that in such scenarios, not all $\mathcal{Z}_n$ subgroups can protect the dark matter stability. To this end we gave several examples where  the dark matter stability is lost as it is unprotected by the residual  $\mathcal{Z}_n$ subgroup. For completeness we also discussed scenarios where a new accidental symmetry might be present in the model, thus protecting the dark matter stability. Finally, we constructed an explicit model with  $ U(1)_{B-L} \to \mathcal{Z}_6$ breaking pattern, where the Dirac nature of neutrinos and dark matter stability are indeed simultaneously protected by the residual $\mathcal{Z}_6$ symmetry.

\begin{acknowledgments}

We thank Ricardo Cepedello and Salvador Centelles Chuli\'a for many useful discussions. RS is supported by the  Spanish grants SEV-2014-0398 and FPA2017-85216-P (AEI/FEDER, UE), PROMETEO/2018/165 (Generalitat Valenciana) and the Spanish Red Consolider MultiDark FPA2017-90566-REDC. The work of C.B. was supported by the Collaborative Research Center SFB1258. EP is supported  by DGAPA-PAPIIT IN107118. 
RS will like to also thank ``El Jefe'' for sponsoring his Formula One GP Mexico tickets.

\end{acknowledgments}

%

\providecommand{\href}[2]{#2}\begingroup\raggedright\endgroup

\end{document}